\documentclass[5p,times,twocolumn]{elsarticle}
\usepackage[T1]{fontenc}
\usepackage{amsmath,amssymb,amsthm}
\usepackage{booktabs,multirow}
\usepackage{url}
\usepackage{xcolor}
\usepackage{algorithm}
\usepackage[noend]{algpseudocode}
\usepackage{graphicx}
\usepackage{tikz}
\usepackage{pgfplots}\pgfplotsset{compat=1.18}
\usepackage[hidelinks]{hyperref}
\usepackage{orcidlink}
\usepackage[capitalise,nameinlink]{cleveref}
\newtheorem{definition}{Definition}
\newtheorem{theorem}{Theorem}
\newtheorem{lemma}{Lemma}

\theoremstyle{remark}\newtheorem{remark}{Remark}
\newcommand{\sys}{\textsc{LinEra}}
\newcommand{\Cl}{\mathsf{Cl}}
\newcommand{\Anc}{\mathsf{Anc}}
\newcommand{\Fk}{F}
\newcommand{\tagsp}{\mathcal{T}}
\newcommand{\AbllamaThreebsessionwriterconvZeroOneCanProp}{10}
\newcommand{\AbllamaThreebsessionwriterconvZeroOneCanaries}{11}
\newcommand{\AbllamaThreebsessionwriterconvZeroOneCanlineraattributed}{1}
\newcommand{\AbllamaThreebsessionwriterconvZeroOneCanperobject}{10}
\newcommand{\AbllamaThreebsessionwriterconvZeroOneCanplainlogical}{11}
\newcommand{\AbllamaThreebsessionwriterconvZeroOneFonenoneZeroZero}{28.5}
\newcommand{\AbllamaThreebsessionwriterconvZeroOneFonesoundZeroZeroOne}{16.1}
\newcommand{\AbllamaThreebsessionwriterconvZeroOneFonesoundrdZeroZeroOne}{26.9}
\newcommand{\AbllamaThreebsessionwriterconvZeroOneerasedZeroZeroOne}{63.2\%}
\newcommand{\AbllamaThreebsessionwriterconvZeroOnemeansourceancestors}{141.9}
\newcommand{\AbllamaThreebsessionwriterconvZeroOnepredictedZeroZeroOne}{70.7\%}
\newcommand{\AbsessionwriterconvZeroOneCanperobject}{7}
\newcommand{\AbsessionwriterconvZeroOneFonenoneZeroZero}{12.9}
\newcommand{\AbsessionwriterconvZeroOneerasedZeroZeroOne}{64.4\%}
\newcommand{\AbwindowEightwriterconvZeroOneFonenoneZeroZero}{20.7}
\newcommand{\AbwindowFourwriterconvZeroOneFonenoneZeroZero}{24.1}
\newcommand{\AbwindowOneSixwriterconvZeroOneFonenoneZeroZero}{18.2}
\newcommand{\AbwindowOneSixwriterconvZeroOneerasedZeroZeroOne}{12.0\%}
\newcommand{\AbwindowTwowriterconvZeroOneCanaries}{11}
\newcommand{\AbwindowTwowriterconvZeroOneCanperobject}{11}
\newcommand{\AbwindowTwowriterconvZeroOneFonenoneZeroZero}{23.5}
\newcommand{\AbwindowTwowriterconvZeroOneerasedZeroZeroOne}{2.0\%}
\newcommand{\Canlineraattributedobjectslost}{56}
\newcommand{\Canlineraattributedtopicleakindescendants}{3}
\newcommand{\CanlinerasoundForgetMs}{5.7}
\newcommand{\CanlinerasoundWriteMs}{0.11}
\newcommand{\Canlinerasoundobjectslost}{526}
\newcommand{\Canperobjectanychannel}{22}
\newcommand{\Canperobjectidentifiedfromvectors}{8}
\newcommand{\Canperobjectrecoveredbydecryption}{14}
\newcommand{\Canperobjecttopicleakindescendants}{22}
\newcommand{\Canplainlogicalanychannel}{35}
\newcommand{\Canplainlogicalidentifiedfromvectors}{26}
\newcommand{\Canplainlogicalrecoveredfromrawbytes}{35}
\newcommand{\Linsoundmeanparents}{24.8}
\newcommand{\Linsoundmeansourceancestors}{167.1}
\newcommand{\Linsoundsingleforgetdescendantsmean}{58.2}
\newcommand{\LmeDerived}{514}
\newcommand{\LmeRadsoundZeroOne}{66.8\%}
\newcommand{\LmeRadsoundZeroZeroOne}{9.9\%}
\newcommand{\LmeSessions}{76}
\newcommand{\LmeSources}{405}
\newcommand{\LocoCanPropExact}{14}
\newcommand{\LocoCanPropTopic}{22}
\newcommand{\LocoCanaries}{35}
\newcommand{\LocoConvs}{5}
\newcommand{\LocoDerived}{967}
\newcommand{\LocoLLMHours}{1.5}
\newcommand{\LocoQuestions}{760}
\newcommand{\LocoSessions}{128}
\newcommand{\LocoSources}{2{,}795}
\newcommand{\LocoTokensPrompt}{124{,}003}
\newcommand{\MemLinAfter}{1}
\newcommand{\MemLinBefore}{11}
\newcommand{\MemLinClosure}{12}
\newcommand{\MemLinSecretsDec}{0}
\newcommand{\MemLinSecretsFiles}{0}
\newcommand{\MemSecrets}{3}
\newcommand{\MemVanBefore}{11}
\newcommand{\MemVanDeleted}{3}
\newcommand{\MemVanFiles}{\texttt{history.db}, \texttt{qdrant/collection/mem/storage.sqlite}}
\newcommand{\QALmeN}{16}
\newcommand{\QALmenoneatZeropZeroSD}{7.4}
\newcommand{\QALoN}{60}
\newcommand{\QALononeatZeropZeroFone}{12.9}
\newcommand{\QALononeatZeropZeroSD}{2.1}
\newcommand{\QALosoundatZeropOneFone}{7.1}
\newcommand{\QALosoundatZeropZeroOneAlive}{95}
\newcommand{\QALosoundatZeropZeroOneFone}{10.5}
\newcommand{\QALosoundrdatZeropOneCalls}{38}
\newcommand{\QALosoundrdatZeropOneFone}{11.7}
\newcommand{\QALosoundrdatZeropOneMin}{20}
\newcommand{\QALosoundrdatZeropZeroOneCalls}{26}
\newcommand{\QALosoundrdatZeropZeroOneFone}{17.2}
\newcommand{\QALosoundrdatZeropZeroOneMin}{14}
\newcommand{\QAWinnoneatZeropZeroFone}{24.1}
\newcommand{\QAWinsoundatZeropOneAlive}{373}
\newcommand{\QAWinsoundatZeropOneFone}{21.3}
\newcommand{\QAWinsoundatZeropOneOrig}{557}
\newcommand{\QAWinsoundatZeropZeroOneFone}{24.1}
\newcommand{\QAWinsoundrdatZeropOneCalls}{67}
\newcommand{\QAWinsoundrdatZeropOneFone}{22.7}
\newcommand{\QAWinsoundrdatZeropOneMin}{11}
\newcommand{\QAWinsoundrdatZeropZeroOneCalls}{8}
\newcommand{\QAWinsoundrdatZeropZeroOneFone}{24.2}
\newcommand{\QSevenLmeQAN}{77}
\newcommand{\QSevenLmeQAnoneatZeropZeroFone}{27.8}
\newcommand{\QSevenLmeQAnoneatZeropZeroSD}{3.6}
\newcommand{\QSevenLmeQAsoundatZeropOneFone}{12.4}
\newcommand{\QSevenLmeQAsoundatZeropZeroOneFone}{26.9}
\newcommand{\QSevenLmeQAsoundrdatZeropOneFone}{29.7}
\newcommand{\QSevenLmeQAsoundrdatZeropZeroOneFone}{28.7}
\newcommand{\QSevenMemLinAfter}{3}
\newcommand{\QSevenMemLinBefore}{130}
\newcommand{\QSevenMemLinClosure}{131}
\newcommand{\QSevenMemLinSecretsDec}{0}
\newcommand{\QSevenMemLinSecretsFiles}{0}
\newcommand{\QSevenMemMessages}{60}
\newcommand{\QSevenMemVanDeleted}{20}
\newcommand{\QSevenMemVanFiles}{\texttt{history.db}}
\newcommand{\QSevenMemVanSecretsFiles}{6}
\newcommand{\QSevenSesCanProp}{60}
\newcommand{\QSevenSesCanaries}{71}
\newcommand{\QSevenSesCanlineraattributedanychannel}{13}
\newcommand{\QSevenSesCanperobjectanychannel}{67}
\newcommand{\QSevenSesCanperobjectrecoveredbydecryption}{60}
\newcommand{\QSevenSesCanplainlogicalanychannel}{70}
\newcommand{\QSevenSesDerived}{2{,}229}
\newcommand{\QSevenSesFanin}{181.5}
\newcommand{\QSevenSesPredZeroZeroOne}{77.2\%}
\newcommand{\QSevenSesQAN}{1{,}323}
\newcommand{\QSevenSesQAnoneatZeropZeroFone}{27.8}
\newcommand{\QSevenSesQAnoneatZeropZeroSD}{0.9}
\newcommand{\QSevenSesQAsoundatZeropOneFone}{2.0}
\newcommand{\QSevenSesQAsoundatZeropZeroOneAlive}{831}
\newcommand{\QSevenSesQAsoundatZeropZeroOneFone}{13.8}
\newcommand{\QSevenSesQAsoundrdatZeropOneCalls}{272}
\newcommand{\QSevenSesQAsoundrdatZeropOneFone}{27.3}
\newcommand{\QSevenSesQAsoundrdatZeropOneMin}{27}
\newcommand{\QSevenSesQAsoundrdatZeropZeroOneCalls}{169}
\newcommand{\QSevenSesQAsoundrdatZeropZeroOneFone}{27.9}
\newcommand{\QSevenSesQAsoundrdatZeropZeroOneMin}{19}
\newcommand{\QSevenSesRadsoundZeroZeroOne}{73.0\%}
\newcommand{\QSevenSesSessions}{272}
\newcommand{\QSevenSesSources}{5{,}953}
\newcommand{\QSevenWinCanProp}{68}
\newcommand{\QSevenWinFanin}{3.8}
\newcommand{\QSevenWinPredZeroZeroOne}{3.8\%}
\newcommand{\QSevenWinQAnoneatZeropZeroFone}{31.8}
\newcommand{\QSevenWinQAsoundatZeropOneFone}{26.6}
\newcommand{\QSevenWinQAsoundatZeropZeroOneFone}{30.7}
\newcommand{\QSevenWinQAsoundrdatZeropOneCalls}{498}
\newcommand{\QSevenWinQAsoundrdatZeropOneFone}{31.9}
\newcommand{\QSevenWinQAsoundrdatZeropZeroOneCalls}{57}
\newcommand{\QSevenWinQAsoundrdatZeropZeroOneFone}{31.8}
\newcommand{\QSevenWinRadsoundZeroOne}{32.7\%}
\newcommand{\QSevenWinRadsoundZeroZeroOne}{3.8\%}
\newcommand{\RadflatZeroZeroOne}{59.7\%}
\newcommand{\RadsoundZeroZeroOne}{74.8\%}
\newcommand{\RetpZeroOnesoundhybrid}{68.3\%}
\newcommand{\RetpZeroZeroOnesoundfactsonly}{18.9\%}
\newcommand{\RetpZeroZeroOnesoundhybrid}{70.2\%}
\newcommand{\RetpZeroZerononefactsonly}{63.2\%}
\newcommand{\RetpZeroZerononehybrid}{79.5\%}
\newcommand{\RrsessionwriterconvZeroOnesounderasedZeroOne}{100.0\%}
\newcommand{\RrsessionwriterconvZeroOnesounderasedZeroZeroOne}{64.4\%}
\newcommand{\RrsessionwriterconvZeroOnesoundmeansourceancestors}{126.2}
\newcommand{\RrsessionwriterconvZeroOnesoundpredictedZeroZeroOne}{67.3\%}
\newcommand{\RrwindowFourwriterconvZeroOnellmcalls}{214}
\newcommand{\RrwindowFourwriterconvZeroOnesounderasedZeroOne}{33.0\%}
\newcommand{\RrwindowFourwriterconvZeroOnesounderasedZeroZeroOne}{3.9\%}
\newcommand{\RrwindowFourwriterconvZeroOnesoundmeansourceancestors}{3.8}
\newcommand{\RrwindowFourwriterconvZeroOnesoundpredictedZeroZeroOne}{3.7\%}
\newcommand{\ScchainedOneZeroZeroZeroZeroZeroclosuredescendantsmedian}{13{,}023}
\newcommand{\ScchainedOneZeroZeroZeroZeroZeroforgetmsmedian}{59.6}
\newcommand{\ScflatOneZeroZeroZeroZeroZeroZeroclosuredescendantsmedian}{9}
\newcommand{\ScflatOneZeroZeroZeroZeroZeroZeroforgetmsmedian}{9.05}
\newcommand{\ScflatOneZeroZeroZeroZeroZeroZerogetmscold}{0.16}
\newcommand{\ScflatOneZeroZeroZeroZeroZeroZeropprfstateKiBafterforgets}{95.4}
\newcommand{\ScflatOneZeroZeroZeroZeroZeroZerorotates}{59.9}
\newcommand{\ScflatOneZeroZeroZeroZeroZeroZerosearchwarmms}{326}
\newcommand{\ScflatOneZeroZeroZeroZeroZeroZerowritemsperobject}{0.28}
\newcommand{\WinCanPropExact}{9}
\newcommand{\WinCanaries}{11}
\newcommand{\WinCanlinerasoundobjectslost}{40}
\newcommand{\WinCanperobjectanychannel}{10}
\newcommand{\WinCanplainlogicalanychannel}{11}

\begin{document}
\begin{frontmatter}
\title{Forget the Lineage: Forward-Secure, Lineage-Complete Cryptographic Erasure for LLM-Agent Memory}
\author[efrei,cnam]{Yulliwas Ameur\corref{cor1}\orcidlink{0000-0003-1435-2982}}
\ead{yulliwas.ameur@efrei.fr}
\cortext[cor1]{Corresponding author.}
\author[cnam]{Billal Mokhtari}
\author[cnam]{Samia Bouzefrane}
\affiliation[efrei]{organization={Efrei Research Lab, Université Paris-Panthéon-Assas},addressline={30--32 avenue de la République},postcode={94800},city={Villejuif},country={France}}
\affiliation[cnam]{organization={Cédric laboratory, Conservatoire national des arts et métiers (Cnam)},addressline={292 rue Saint-Martin},postcode={75003},city={Paris},country={France}}

\begin{abstract}
Large language model (LLM) agents keep a long-term memory that the model itself writes: summaries, extracted facts and embeddings derived
from what users said. Deleting a message is therefore not enough: its derivatives, storage residue
and backups survive. We define \emph{lineage-complete forward-secure erasure}: after a forget
request, an adversary that later obtains all storage and the agent's current keys learns nothing about the
forgotten input or any memory derived from it. We present \sys{}, which encrypts each memory under a key derived
from its parents' keys along the derivation graph; source keys come from a puncturable pseudorandom function, so forgetting is a
single puncture and every descendant becomes unreadable without being located. A \emph{sound} lineage rule records
as parents everything the writer saw, which confines all influence of a source to its descendants. We prove
security and lineage integrity and integrate \sys{} with mem0. With real LLMs from two families (1.5B to 7B
parameters) writing the memory of LoCoMo conversations, an attacker holding every file and the current keys
recovers, on all ten conversations written by the largest model, \QSevenSesCanplainlogicalanychannel{} of
\QSevenSesCanaries{} forgotten secrets after logical deletion, \QSevenSesCanperobjectanychannel{} with per-object or
epoch keys, and none with \sys{}; the same holds for every writer tested. The price is a larger forgetting radius,
which follows a simple law in the writer's fan-in: forgetting 1\% of turns erases \QSevenSesRadsoundZeroZeroOne{}
of derived memories with session-level summaries but only \QSevenWinRadsoundZeroZeroOne{} with four-turn windows,
which also answer better; re-deriving restores what is lost.
\end{abstract}

\end{frontmatter}

\section{Introduction}\label{sec:intro}
Large language model (LLM) agents now keep a long-term memory of their users. Memory layers such
as mem0 \cite{chhikara2025mem0} turn conversations into facts and summaries, store them with their embeddings, and
bring them back into later prompts. This memory is personal data, and users can ask to have it erased
\cite{gdpr2016}. Erasure is harder than it looks, because most of the memory was not written by the user: it was
written by the model, from what the user said.

\emph{Example.} Alice tells her assistant that her new door code is 74200. That evening the agent writes a
summary of the day and a fact ``Alice's door code is 74200''. A week later it writes a profile entry from that
summary. When Alice asks the agent to forget her message, deleting the message is not enough: the fact, the summary
and the profile entry still hold the code, and so may the free pages of the database, the vector index and last
night's backup. In our experiments, the 1.5-billion-parameter model that wrote the memory copied the exact secret
into a derived memory for \LocoCanPropExact{} of \LocoCanaries{} planted secrets, and its topic for
\LocoCanPropTopic{} of them.

This paper asks a precise question: \emph{after a forget request, can we guarantee that the forgotten input and
everything derived from it are unreadable, for good, even for someone who later steals the whole storage and the
system's current keys?} Existing answers fall short. Logical deletion across memory layers
\cite{singhal2026forgetagent} trusts the store after the fact. Replaying the agent without the input
\cite{yao2026execstate} is costly and leaves storage residue. Cryptographic erasure with one key per object, as
in Holepunch \cite{ratliff2024holepunch}, or one key per epoch, as proposed against soft-deleted vectors
\cite{chakraborttii2026ghost}, erases the targeted object but not the memories that the model derived from it.

\paragraph{Our approach}
We make the key of every memory depend on the keys of the memories it was derived from. Source keys come from a
puncturable pseudorandom function (PPRF) \cite{boneh2013constrained,ggm1986}; the key of a derived memory is an
HKDF of its parents' keys. Forgetting a source is a single \emph{puncture} of the PPRF. Every descendant then
becomes undecryptable at once, wherever its ciphertext lives, without being located. To know the parents, we use a
\emph{sound} lineage rule: the parents of a memory are everything the model saw when it wrote it. This rule needs
no access to the model, only to the prompt, and it is what makes the cryptographic guarantee meaningful
(\cref{lem:confine}).

\paragraph{Findings}
On \LocoSessions{} LoCoMo sessions whose memory was written by a real model, an attacker holding every file and
the current keys recovers \Canplainlogicalanychannel{} of \LocoCanaries{} forgotten secrets after logical
deletion, \Canperobjectanychannel{} after per-object or epoch keys, and none after \sys{} with the sound rule.
The guarantee has a price, and we measure it precisely: a memory written from $f$ source turns survives a random
forget rate $p$ with probability close to $(1-p)^{f}$. A typical session-level writer that also reads earlier
summaries has a fan-in of more than a hundred turns, so forgetting 1\% of the turns erases \RadsoundZeroZeroOne{}
of the derived memories. The remedy is to keep the writer's context small: when the same model extracts facts from four-turn windows,
the radius at 1\% falls to \RrwindowFourwriterconvZeroOnesounderasedZeroZeroOne{} (against
\RrsessionwriterconvZeroOnesounderasedZeroZeroOne{} for the session writer on the same conversations), and answer
quality is not hurt. Re-deriving the lost memories from what survives restores answer quality in both settings, at the cost of one
model call per affected session or window. A larger writer, Qwen2.5-7B, on all ten LoCoMo conversations confirms
each of these findings: \QSevenSesCanperobjectanychannel{} of \QSevenSesCanaries{} forgotten secrets leak with
per-object keys and none with \sys{}, and the radius at 1\% is \QSevenSesRadsoundZeroZeroOne{} for the session writer
(predicted \QSevenSesPredZeroZeroOne{}) and \QSevenWinRadsoundZeroZeroOne{} for four-turn windows.

\paragraph{Contributions}
\begin{enumerate}
\item A security definition, LC-FS, for lineage-complete forward-secure erasure in agent memory, and a lemma that
connects it to what an LLM writer can have seen (\cref{sec:model}).
\item \sys{}, a key graph over a GGM puncturable PRF with an encrypted vector index whose shard keys follow the
lineage, crash-consistent state and compaction, and a 140-line integration into mem0 (\cref{sec:construction}).
\item Proofs of LC-FS security in the random-oracle model and, with a polynomial loss, in the standard model, and of lineage integrity (\cref{sec:security}, Appendices~\ref{app:proof} and~\ref{app:std}).
\item An evaluation with real LLM writers from two model families and three sizes (1.5B to 7B parameters) on LoCoMo
and LongMemEval: four leakage channels,
the fan-in law of the forgetting radius (checked for writer contexts from 2 turns to whole sessions), its utility
cost and two remedies (re-derivation and small-context writing), system costs up to $10^6$ objects, and a case study
on mem0 (\cref{sec:eval}).
\end{enumerate}
\section{Background: How Agent Memories Grow}\label{sec:background}
Long-term memory layers for LLM agents, such as mem0 \cite{chhikara2025mem0}, follow the same loop. New messages
arrive. The framework retrieves related memories by embedding search and shows them, together with the new
messages and a short window of recent ones, to a language model. The model writes new memories: extracted facts
(``Alice moved to Lyon''), updates of older facts, or session summaries. Each memory is embedded and indexed. Later
retrievals bring these memories back into new prompts, so memories are written from memories.

Two consequences matter for erasure. First, most of what the store holds was \emph{written by the model}, not
received from the user. In our LoCoMo run, the writer produced \LocoDerived{} derived memories from
\LocoSources{} turns. Second, information moves between objects in ways that the store does not see: a fact can
paraphrase a message, and a summary can carry a detail from an older summary. Deleting the message that a user
points at leaves these copies in place. Storage adds a third layer of copies: deleted rows stay in free database
pages, soft-deleted vectors stay in index files \cite{chakraborttii2026ghost}, and backups keep everything. As
we found when reading its code, mem0 2.2.1 even writes the old text of a deleted memory into its history table.
\section{Related Work}\label{sec:related}

\paragraph{Puncturable primitives and forward security}
Puncturable PRFs were introduced as constrained PRFs \cite{boneh2013constrained,kiayias2013delegatable,boyle2014functional},
with GGM \cite{ggm1986} as the basic instance and adaptive security studied in \cite{fuchsbauer2014adaptive}.
Puncturable encryption gives forward secrecy to asynchronous messaging \cite{green2015puncturable} and to 0-RTT
key exchange through Bloom-filter encryption \cite{derler2021bfe} and session resumption \cite{aviram2021resumption}.
Puncturable key wrapping \cite{backendal2022pkw} punctures a key-encryption key over a fixed set of tags. All of
these treat the protected items as independent. Our contribution is to make the key of an item depend on the keys
of the items it was \emph{derived from}, so that one puncture erases a closure that is not known in advance.

\paragraph{Secure and cryptographic deletion}
Reardon et al.\ survey secure deletion \cite{reardon2013sok}; cryptographic erasure (delete the key, not the data)
is the standard answer when storage cannot be overwritten. Holepunch \cite{ratliff2024holepunch} deletes files with
a GGM PPRF and crash-consistent state, and we reuse its state-handling ideas. Cloud schemes provide fine-grained
erasure in outsourced storage \cite{wei2023fsdabpe} and erasable, portable storage with oblivious key management
\cite{yang2026epcs}. None of them models data that is \emph{derived} from deleted data by a third process, here a
language model.

\paragraph{Forgetting in LLM agents}
Machine unlearning removes the influence of training data from model weights \cite{bourtoule2021unlearning}; it is
complementary to our work, which concerns external memory. For agents, ForgetAgent \cite{singhal2026forgetagent}
deletes logically across the layers of a multi-layer memory and issues deletion receipts; execution-state unlearning
\cite{yao2026execstate} replays an agent's execution without the forgotten input; benchmark studies \cite{lin2026aubench} and behavioural analyses \cite{liu2026deletion,mukkuzhi2026ghostechoes} show that deleted
memories keep influencing later answers. Ghost Vectors \cite{chakraborttii2026ghost} shows that soft-deleted
vectors in HNSW indexes can be reconstructed and proposes per-epoch encryption. MemLineage \cite{ouyang2026memlineage}
records a signed derivation DAG to fight memory poisoning, i.e., for integrity rather than erasure. MemoRepair
\cite{zhao2026memorepair} and agentic unlearning \cite{wang2026agenticunlearning} withdraw or invalidate the memories
derived from a deleted source, logically and without a guarantee against a later theft of the storage, and a recent
benchmark \cite{xie2026bemula} finds that current memory systems rarely complete a forget request end to end. These works
either rely on the store behaving honestly after deletion, or erase only the targeted object. \sys{} is, to our
knowledge, the first scheme with a formal guarantee that covers derived memories and holds against a later theft of
the storage and of the current secret state.

\paragraph{Revocation in agent memory}
In earlier work with colleagues \cite{ameur2026recovery} we bounded the \emph{work} needed to repair an agent memory
when a source is found to be wrong (an integrity and availability question) using the same dependency graph.
\sys{} addresses the confidentiality side: once a source must be forgotten, how to make it and all its
derivatives unreadable for good. The two can share the lineage capture of \cref{sec:construction}.
\section{System and Threat Model}\label{sec:model}

\subsection{Memory objects and their lineage}
An agent memory is a set of \emph{objects}. A \emph{source} object is an input that the agent received: a user
message, a tool result, a document chunk. A \emph{derived} object is text or a vector that the agent itself wrote
from other objects: a session summary, an extracted fact, a merged profile entry, a note, an embedding.
Each object $o$ has a public random tag $t_o\in\tagsp=\{0,1\}^{n}$ ($n=32$ in our implementation) and a set of
parents $P(o)$ of earlier objects. Sources have no parents. The parent relation forms a directed acyclic graph
(DAG), which we call the \emph{lineage graph}. We write $\Anc(o)$ for the ancestors of $o$ and, for a set
$\Phi$ of objects, $\Cl(\Phi)=\Phi\cup\{o:\Anc(o)\cap\Phi\neq\emptyset\}$ for its \emph{lineage closure}:
$\Phi$ and everything derived from it, directly or not.

\emph{Example.} On Monday Alice tells the agent ``my new door code is 74200'' (source $s$). That evening the agent
writes a summary $r$ of the day (parents: all Monday messages) and a fact $f$ ``Alice has a new door code''
(same parents). On Tuesday it retrieves $r$ to write a weekly profile $w$. Then $\Cl(\{s\})=\{s,r,f,w\}$.

\subsection{Which parents? The lineage rule}
The lineage graph is not given; the memory pipeline must record it. A \emph{lineage rule} decides $P(d)$ when a
derived object $d$ is written. We study two rules.
\begin{itemize}
\item \textbf{Sound rule.} $P(d)$ is every object whose plaintext is in the writer's context when $d$ is written:
the new messages, the conversation window that the framework shows, and every memory returned by retrieval.
\item \textbf{Attributed rule.} $P(d)$ is a subset chosen by an attribution function; in our experiments, the
three context objects whose embeddings are closest to $d$.
\end{itemize}
The sound rule needs no knowledge of the model. It only observes what enters the prompt. Its justification is
the following simple lemma, which is the bridge between information flow and cryptography.

\begin{lemma}[Context confinement]\label{lem:confine}
Model the writer as a function $W(\mathit{ctx},\theta,\omega)$ of its context, its weights $\theta$ and its
randomness $\omega$. Let the sources be independent random variables and record lineage with the sound rule.
Then, for every source $s$, the joint content of all objects outside $\Cl(\{s\})$ is a function of the other
sources, $\theta$ and the writer randomness; hence it is independent of the content of $s$.
\end{lemma}
\begin{proof}
By induction in creation order. A source $o\neq s$ is independent of $s$ by assumption. A derived $d\notin\Cl(\{s\})$
has no parent in $\Cl(\{s\})$, since otherwise $d$ would be in the closure. By the sound rule its context is
exactly the content of its parents, which by induction does not depend on $s$; so neither does $d=W(\cdot)$.
\end{proof}
The lemma is the reason why the sound rule is ``sound'': whatever a forgotten source could have influenced is
inside its closure. It also states the three limits that no storage scheme can remove: what the weights $\theta$
already know, information about $s$ that other sources also carry (if Alice repeats her code in another message,
forgetting one message does not forget the code), and anything that was copied out of the system before the
request. The attributed rule gives no such guarantee: a fact can copy a detail from a context object that the
attribution function did not select.

\subsection{Adversary}
The adversary sees the storage at all times: ciphertexts, the lineage graph, files, backups, free pages and
logs. This captures an honest-but-curious cloud host or a backup operator. It does \emph{not} see the secret
state $S$ of the store before the forget requests. At some later time it \emph{corrupts} the system and obtains
everything: all storage ever written and the \emph{current} secret state. This captures a stolen laptop, a
seized server, a compelled disclosure or a compromised operator after the erasure request. The goal is that
nothing about forgotten objects and their descendants can be learned from this view. RAM content at the very
moment of forgetting, side channels on the running process and the model weights are out of scope.

\subsection{Security definition}
\begin{definition}[LC-FS security]\label{def:lcfs}
Let $\Pi=(\mathsf{Setup},\mathsf{Put},\mathsf{Forget})$ be a memory store. The game
$\mathbf{G}^{\mathrm{lcfs}}_{\Pi}(\mathcal{A})$ runs as follows.
\begin{enumerate}
\item The challenger runs $S\gets\mathsf{Setup}(1^\lambda)$ and draws $b\gets\{0,1\}$. The public view
$\Sigma$ (all ciphertexts ever written, including overwritten and deleted ones, and the lineage graph) starts empty.
\item $\mathcal{A}$ adaptively calls three oracles and always receives the new part of $\Sigma$:
  \begin{itemize}
  \item $\textsc{New}(P,m)$: stores content $m$ as a new object with parents $P$ (existing objects) and returns its tag;
  \item $\textsc{Chal}(P,m_0,m_1)$, with $|m_0|=|m_1|$: stores $m_b$ as a new object with parents $P$; the object is
        marked as a challenge object;
  \item $\textsc{Forget}(t)$: runs $\mathsf{Forget}(S,t)$ and adds $t$ to $\Phi$.
  \end{itemize}
\item $\mathcal{A}$ calls $\textsc{Corrupt}$ and receives the current state $S$. It outputs $b'$.
\end{enumerate}
$\mathcal{A}$ is \emph{valid} if every challenge object is in $\Cl(\Phi)$ at corruption time. Its advantage is
$\mathrm{Adv}^{\mathrm{lcfs}}_{\Pi}(\mathcal{A})=|2\Pr[b'=b]-1|$. $\Pi$ is \emph{lineage-complete forward-secure}
(LC-FS) if this advantage is negligible for every efficient valid $\mathcal{A}$.
\end{definition}
Three design choices of the definition deserve a word. First, the adversary chooses the contents and parents of
all objects, so the definition covers any writer, honest or adversarial. Second, several challenge objects are
allowed, anywhere in the closure. This models a summary of a forgotten message, a fact extracted from that
summary, and so on: by \cref{lem:confine} these are exactly the objects whose content may depend on the
forgotten source, and the game lets them depend on the bit $b$. Third, the adversary gets every ciphertext ever
written, so backups and storage residue are included by construction. Note that a store which deletes rows but
keeps keys (logical deletion), or that erases only the forgotten object (per-object keys), is trivially broken:
the adversary forgets a source and wins with a challenge child.
\section{The \sys{} Construction}\label{sec:construction}

\subsection{Building block: a puncturable PRF}
A puncturable pseudorandom function (PPRF) \cite{boneh2013constrained,kiayias2013delegatable,boyle2014functional}
is a PRF $\Fk(K,\cdot)$ with an extra operation $\mathsf{Puncture}(K,x)$. It returns a new key $K_{\{x\}}$ that still
computes $\Fk(K,y)$ for every $y\neq x$, but reveals nothing about $\Fk(K,x)$. We use the classic tree construction of
Goldreich, Goldwasser and Micali (GGM) \cite{ggm1986} on $n$-bit inputs, with the length-doubling generator
$G(s)=(G_0(s),G_1(s))$, $G_b(s)=\mathrm{SHA\text{-}256}(b\|s)$. The key is a set of tree nodes (prefix, seed) that
together cover all non-punctured inputs; initially it is the root. To evaluate $x$, one walks down from the node
that covers $x$. To puncture $x$, one removes that node and keeps the siblings of the path to $x$: at most $n$
new nodes. Holepunch \cite{ratliff2024holepunch} uses the same tool to delete files from a disk; we use it as the
root of a key graph.

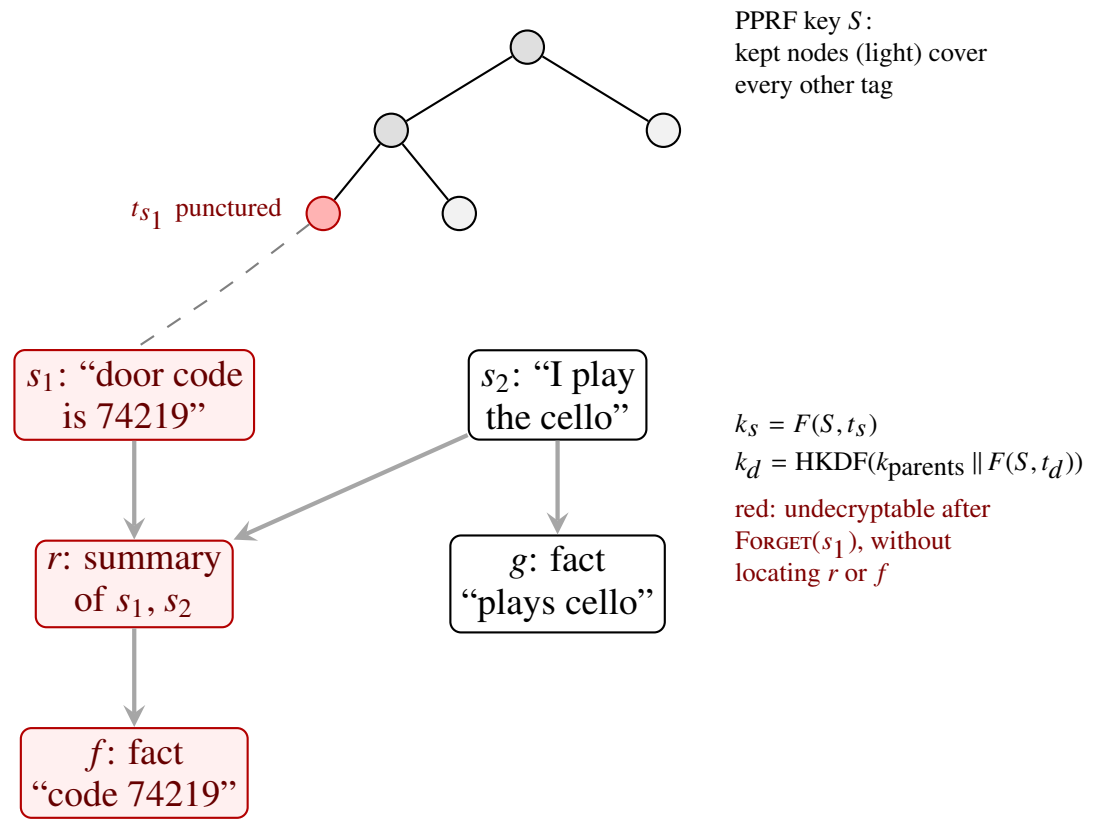
\begin{figure}[t]\centering
\begin{tikzpicture}[font=\scriptsize, >=stealth,
  obj/.style={draw, rounded corners=2pt, align=center, minimum height=5.2mm, inner sep=2pt, fill=white},
  dead/.style={obj, draw=red!70!black, fill=red!6, text=red!40!black},
  keyarrow/.style={->, thick, gray!70},
  tree/.style={draw, circle, inner sep=0.8pt, minimum size=2.2mm}]
\node[tree, fill=gray!25] (root) at (0,0) {};
\node[tree, fill=gray!25] (a) at (-0.9,-0.55) {};
\node[tree, fill=gray!10] (b) at (0.9,-0.55) {};
\node[tree, fill=red!30, draw=red!70!black] (ls1) at (-1.35,-1.1) {};
\node[tree, fill=gray!10] (ls1s) at (-0.45,-1.1) {};
\draw (root)--(a) (root)--(b) (a)--(ls1) (a)--(ls1s);
\node[font=\tiny, text=red!50!black, anchor=east] at (-1.5,-1.1) {$t_{s_1}$ punctured};
\node[font=\tiny, align=left, anchor=west] at (1.25,-0.05) {PPRF key $S$:\\kept nodes (light) cover\\every other tag};
\node[dead] (s1) at (-2.6,-2.3) {$s_1$: ``door code\\ is 74219''};
\node[obj] (s2) at (0.2,-2.3) {$s_2$: ``I play\\ the cello''};
\node[dead] (r) at (-2.6,-3.55) {$r$: summary\\ of $s_1,s_2$};
\node[obj] (g) at (0.2,-3.55) {$g$: fact\\ ``plays cello''};
\node[dead] (f) at (-2.6,-4.8) {$f$: fact\\ ``code 74219''};
\draw[keyarrow] (s1)--(r); \draw[keyarrow] (s2)--(r); \draw[keyarrow] (s2)--(g); \draw[keyarrow] (r)--(f);
\draw[dashed, gray] (ls1) .. controls (-2.0,-1.6) .. (s1.north);
\node[align=left, anchor=west, font=\tiny] at (1.25,-3.0) {$k_{s}=F(S,t_s)$\\[1pt]$k_d=\mathrm{HKDF}(k_{\text{parents}}\,\|\,F(S,t_d))$\\[3pt]\textcolor{red!50!black}{red: undecryptable after}\\\textcolor{red!50!black}{\textsc{Forget}$(s_1)$, without}\\\textcolor{red!50!black}{locating $r$ or $f$}};
\end{tikzpicture}
\caption{\sys{} in one example. Keys flow along the lineage (grey arrows). Forgetting $s_1$ punctures the PPRF at
$t_{s_1}$: $k_{s_1}$ can no longer be computed, so neither can the keys of its descendants $r$ and $f$, even by an
attacker who later steals the store and the current key $S$. $s_2$ and $g$ stay readable.}\label{fig:linera}
\end{figure}
\subsection{Key derivation along the lineage graph}
The secret state is a PPRF key $S$. Every object $o$ has a key
\begin{equation}\label{eq:key}
\begin{aligned}
k_o&=\Fk(S,t_o) &&\text{if }P(o)=\emptyset,\\
k_o&=\mathrm{HKDF}\big(k_{p_1}\|\cdots\|k_{p_m}\|\Fk(S,t_o);\ I_o\big) &&\text{otherwise,}
\end{aligned}
\end{equation}
where $I_o=\texttt{"lf/lineage"}\|t_o$, $p_1<\dots<p_m$ are the parents sorted by tag, and HKDF is used with an empty salt \cite{krawczyk2010hkdf,rfc5869}.
The content $m$ of $o$ is stored as $c_o=\mathrm{AEAD}_{k_o}(m;\ \mathrm{AD}=t_o\|t_{p_1}\|\cdots\|t_{p_m})$ with
ChaCha20-Poly1305 and a random nonce \cite{rfc8439,rogaway2002aead}. The tags and the parent lists are stored in
clear next to the ciphertext. Two properties follow at once.
\begin{itemize}
\item \emph{One puncture erases a whole closure.} To compute $k_d$ one needs the keys of all parents, hence, by
induction, $\Fk(S,t_a)$ for every ancestor $a$. After $\mathsf{Puncture}(S,t_s)$, no descendant of $s$ can be decrypted,
and the store never has to \emph{find} them.
\item \emph{Every object can be forgotten alone.} The term $\Fk(S,t_o)$ in the derived case gives each derived
object its own puncture point. Forgetting a wrong fact punctures the fact and its descendants only.
\end{itemize}
Including the parents in the associated data binds each ciphertext to its lineage: a storage attacker who rewires
$P(d)$ makes decryption fail (integrity, \cref{sec:security}).

\begin{algorithm}[t]\caption{\sys{} operations (state $S$, store $\Sigma$, key cache $C$).}\label{alg:ops}
\small
\begin{algorithmic}[1]
\Procedure{Put}{$m,P$}
  \State $t\gets$ fresh random tag; record $(t,P)$
  \State $k\gets$ \Call{Key}{$t$}; $c\gets\mathrm{AEAD}_k(m;\ t\|\mathrm{sort}(P))$; append $(t,P,c)$ to $\Sigma$
  \State \Return $t$
\EndProcedure
\Procedure{Key}{$t$} \Comment{memoised in $C$}
  \State $y\gets \Fk(S,t)$ \Comment{fails if $t$ was punctured}
  \If{$P(t)=\emptyset$} \Return $y$ \EndIf
  \State \Return $\mathrm{HKDF}(k_{p_1}\|\cdots\|k_{p_m}\|y;\ I_t)$ \Comment{$k_p=$\,\Call{Key}{$p$}}
\EndProcedure
\Procedure{Forget}{$t$}
  \State $D\gets\{t\}\cup\mathrm{desc}(t)$; decrypt the vector shards that contain $D$
  \State $S\gets\mathsf{Puncture}(S,t)$; write $S$ atomically (temp file, fsync, rename)
  \State drop $C[u]$ for $u\in D$; rebuild the shards without $D$ (new version)
  \State delete the rows of $D$ \Comment{optional; security does not rely on it}
\EndProcedure
\Procedure{Rotate}{} \Comment{compaction}
  \State $S'\gets$ fresh PPRF key; re-encrypt every readable object under the keys derived from $S'$; $S\gets S'$
\EndProcedure
\end{algorithmic}
\end{algorithm}

\subsection{Forgetting, crash consistency and compaction}
\Cref{alg:ops} summarises the operations. A forget request is durable as soon as the new state is on disk; we write it
with the usual temporary-file, \texttt{fsync}, rename sequence, so a crash leaves either the old or the new state.
Deleting rows afterwards is only hygiene. Each puncture adds at most $n$ nodes of 40~bytes to $S$. \textsc{Rotate}
re-encrypts the survivors under a fresh PPRF key and brings $S$ back to one node; it can run in the background,
like the compaction of Holepunch \cite{ratliff2024holepunch}.

\subsection{An encrypted vector index whose keys follow the lineage}
Agent memories are found by embedding search. Embeddings are themselves derived data: text can be largely
recovered from them \cite{morris2023inversion}, and soft-deleted vectors survive in popular index files
\cite{chakraborttii2026ghost}. \sys{} therefore keeps vectors in \emph{shards} of at most $B$ objects (we use
$B=64$ in the agent experiments and $B=256$ for scaling). A shard with members $M$ and version $v$ is one AEAD
ciphertext under
\[k_{\text{shard}}=\mathrm{HKDF}\big(\|_{u\in\mathrm{sort}(M)}k_u;\ \texttt{"lf/shard"}\|\mathit{sid}\|v\big).\]
A shard can be decrypted only while \emph{all} its members are readable. When a forget request hits a member, the
shard is decrypted first, the PPRF is punctured, and the shard is rebuilt without the closure under version
$v+1$. Old copies of the shard, in backups or free pages, die with the punctured key. Search decrypts and caches
the shards in RAM and scans them; an approximate graph index \cite{malkov2020hnsw} can be built per shard inside the
same envelope. If the process crashes between the puncture and the rebuild, the surviving members are simply
re-embedded from their (still readable) text.

\subsection{Capturing lineage in a real memory framework}
The sound rule needs one hook: the store must know what the writer sees. We implemented it for mem0
\cite{chhikara2025mem0} (version 2.2.1), an open-source memory layer, in about 140 lines. The
\sys{} vector store implements mem0's storage interface. A thin wrapper stores each incoming message as a source
object and opens a \emph{lineage context}. During the call, every memory that mem0 retrieves through the store's
\texttt{search} or \texttt{get} is added to the context; every memory that mem0 then inserts or updates receives
the context as parents. The context also contains the earlier messages that mem0 shows the model (its last ten
messages) when, and only when, the model is called. We found that mem0 keeps a second, plaintext SQLite database
of raw messages and of the history of each memory; its delete operation even \emph{writes} the old text of the
deleted memory into it. The wrapper therefore keeps that history in RAM only.

\subsection{Two ways to pay less for the guarantee}
The sound rule erases more than what logical deletion erases: it also erases memories that merely \emph{saw} the
forgotten source. \Cref{sec:eval} measures this \emph{forgetting radius}. We propose two remedies that keep the
guarantee.
\begin{itemize}
\item \textbf{Re-derivation.} After a forget request, every session that lost derived memories is summarised again
by the same writer, from its surviving sources and surviving memories. The new memories get sound parents.
The cost is one model call per affected session.
\item \textbf{Small-context writing.} The radius grows with the number of source turns that the writer sees,
directly or through retrieved summaries. Writing memories from short windows of a few turns, without feeding
earlier summaries back into the writer, keeps every memory's parent set small, so a forgotten source reaches
only the few memories written from its window.
\end{itemize}
\section{Security Analysis}\label{sec:security}

We model $G_0,G_1$ (SHA-256 with a one-byte prefix) and HKDF as independent random oracles \cite{bellare1993ro}
and the AEAD as one-time IND-CPA secure. Let $\lambda=256$ be the seed and key length.

\begin{theorem}[\sys{} is LC-FS secure]\label{thm:main}
Let $\mathcal{A}$ be a valid adversary that makes $q_G$ queries to $G$ and $q_H$ queries to HKDF, forgets a set
$\Phi$ of objects, and causes $N_K$ keys (object keys and shard keys) to be derived. Let $C$ be the number of
ciphertexts in $\Sigma$ that encrypt challenge content (challenge objects and the shard versions that contain
their vectors). Then there is an adversary $\mathcal{B}$, running in about the time of $\mathcal{A}$, such that
\begin{multline*}
\mathrm{Adv}^{\mathrm{lcfs}}_{\sys}(\mathcal{A})\ \le\ \frac{2\,q_G\,|\Phi|\,(n+1)+2\,q_H\,N_K}{2^{\lambda}}\\
+\ 2C\cdot\mathrm{Adv}^{\mathrm{ot\text{-}ind}}_{\mathrm{AEAD}}(\mathcal{B}).
\end{multline*}
\end{theorem}

\emph{Proof idea} (full proof in Appendix~\ref{app:proof}). Three game hops. (i) After puncturing, the GGM nodes on
the paths to the punctured tags are never revealed; in the random-oracle model their values, and so the leaf values
$\Fk(S,t)$ for $t\in\Phi$, look uniform unless $\mathcal{A}$ queries the oracle on one of at most $|\Phi|(n+1)$ hidden
256-bit seeds. (ii) Every key in the closure is then an HKDF output on an input that contains at least one uniform,
hidden 256-bit block: its own punctured leaf, or the key of a parent in the closure. By induction along the DAG,
all these keys look uniform. Shard keys are handled the same way, since a shard that contains a closure member
mixes that member's key. (iii) Each challenge ciphertext is then produced under a fresh uniform key used once, and
one-time IND-CPA security of the AEAD hides $b$. With $q_G=q_H=2^{64}$, $|\Phi|=N_K=2^{30}$ and $n=32$, the
first term is below $2^{-155}$.

\begin{remark}[Standard model]
The random oracles can be removed. Appendix~\ref{app:std} proves LC-FS from the PRG security of $G$, a
block-keyed PRF assumption on HKDF (a dual-PRF-style property of HMAC \cite{bellare2015hmac,krawczyk2010hkdf})
and one-time IND-CPA of the AEAD, with key separation between encryption and derivation. GGM is only selectively
secure \cite{boneh2013constrained,boyle2014functional,kiayias2013delegatable}, and our adversary chooses which
objects to forget adaptively; we avoid the exponential cost of guessing all punctured points
\cite{fuchsbauer2014adaptive} by guessing only one forgotten ancestor of each challenge object, so the loss is
polynomial (a factor $C'N\ell$, \cref{thm:std}).
\end{remark}

\begin{theorem}[Lineage integrity]\label{thm:int}
Without the secret state, an adversary that modifies the storage cannot make \textsc{Get}$(t)$ return a content
other than the one stored for $t$, nor make the store accept a parent list other than the one recorded at
creation, except with probability $\mathrm{Adv}^{\mathrm{int\text{-}ctxt}}_{\mathrm{AEAD}}$ per forged ciphertext.
\end{theorem}
\begin{proof}[Proof sketch]
The key of $t$ depends on its recorded parents through \eqref{eq:key}, and the parent tags are part of the
associated data. Changing the parent list either changes the key (a ciphertext under a key the adversary never
saw) or the associated data; both are ciphertext forgeries. Our test suite checks this case (a rewired summary
fails to decrypt).
\end{proof}

\paragraph{What leaks}
The lineage graph, object sizes and creation times are in clear: the store needs the parent tags to derive keys
before decrypting. After forgetting, the adversary therefore learns that some object with $k$ children existed,
but not its content. Hiding the graph would require encrypting it under a key that is itself forward-secure; we
leave this to future work. The scheme also does not protect the model weights, RAM at the time of forgetting, or
copies made before the request (\cref{lem:confine}).

\paragraph{Why the baselines fail the definition}
\emph{Logical deletion} keeps all keys (or no encryption at all): the adversary reads the challenge child.
\emph{Per-object keys} (Holepunch-style) puncture only the forgotten object: its children keep their own keys.
\emph{Epoch keys} (the fix proposed for soft-deleted vectors \cite{chakraborttii2026ghost}) re-encrypt the other objects of
the forgotten object's epoch under a fresh key, so every child remains readable. Each attack uses one forget and one
challenge child; \cref{sec:eval} shows that real LLM pipelines create such children.
\section{Implementation}\label{sec:impl}
\sys{} is a Python library of about 400 lines. It uses the \texttt{cryptography} package for ChaCha20-Poly1305 and
HKDF-SHA-256 and SQLite for storage. The GGM PPRF works on 32-bit tags and stores each frontier node as a
(depth, prefix, seed) triple; the state file is written with the temporary-file, \texttt{fsync}, rename sequence and
is meant to be sealed by the platform key store. Object keys are memoised in RAM; a forget request drops the cached
keys of its closure. The vector index of \cref{sec:construction} stores each shard as one ciphertext and seals the
open shard on commit. The mem0 adapter takes about 140 lines. Three baselines share the same interface:
\emph{logical deletion} (plaintext rows and an HNSW index \cite{malkov2020hnsw} from \texttt{hnswlib} with soft
deletion, as offered by \texttt{hnswlib}), \emph{per-object keys} (a GGM key per object, no lineage, in the
spirit of Holepunch \cite{ratliff2024holepunch}) and \emph{epoch keys} (one key per epoch of five sessions; a
forget re-encrypts the rest of the epoch and destroys the old key, as proposed against soft-deleted vectors
\cite{chakraborttii2026ghost}). In the two encrypted baselines the vector of an object is encrypted under the same
key as its text. A test suite (12 tests) checks puncturing, serialisation, closure erasure, theft after
forgetting, compaction, lineage tampering, key derivation along a 5{,}000-deep chain with a cold cache (derivation
is iterative, not recursive), shard rebuilds, stale shard backups, and the mem0 adapter.
\section{Evaluation}\label{sec:eval}
We answer four questions. Do forgotten secrets survive in an LLM-written memory under each scheme (E1, E2)? How
far does a forget request reach, and what controls it (E5)? What does the guarantee cost in answer quality (E3)?
What does it cost in time and space (E4)? We end with a case study on mem0.

\subsection{Setup}\label{sec:setup}
\paragraph{Machine and models} The memory writer, the re-deriver and the question answerer are always the same
open model, at temperature~0. The main runs use Qwen2.5-1.5B-Instruct \cite{qwen2025} quantised to 4 bits
(Q4\_K\_M, llama.cpp) on CPU: the memory logs and the system costs come from one virtual machine with two cores of
an Intel Xeon at 2.1~GHz and 7.8~GiB of RAM, and every answer of \cref{tab:qa,tab:ablation} was computed on one
Apple M5 Pro (llama.cpp 0.3.36, CPU only, fresh cache), so that each table comes from a single machine. To check
that our findings do not depend on one model family, we repeat the main runs on two conversations with
Llama-3.2-3B-Instruct \cite{grattafiori2024llama3} (4 bits). An extended campaign (\cref{sec:e7b}) replays
everything with a larger writer, Qwen2.5-7B-Instruct in 16-bit precision (vLLM 0.33 \cite{kwon2023vllm}, one NVIDIA H100), on all ten
LoCoMo conversations. Embeddings come from bge-small-en-v1.5 \cite{xiao2023cpack} (384 dimensions). The random seed
is 20261109 everywhere.

\paragraph{Memory pipeline} We replay LoCoMo \cite{maharana2024locomo}, a benchmark of very long two-person
conversations. Because the main writer runs on CPU, it covers the first \LocoConvs{} of the 10 conversations
(the extended campaign covers all of them):
\LocoSessions{} sessions and \LocoSources{} turns. For each session, in order, the pipeline retrieves the three
most similar earlier derived memories and asks the model for a summary and up to eight facts, as mem0-style
extractors do. The model wrote \LocoDerived{} derived memories (\LocoTokensPrompt{} prompt tokens,
\LocoLLMHours{}~hours of generation). We log exactly what each prompt contained. Every scheme and every lineage
rule is then evaluated by replaying this log, so all of them store the very same texts written by the very same
model; only the recorded parents and the cryptography differ.

\paragraph{Planted secrets} Before the run, we inserted one extra turn in a seeded quarter of the sessions:
\LocoCanaries{} \emph{canary} turns such as ``my new apartment door code is 28352, and the building manager is
called Thalwyndor'', built from four templates with a random five-digit number and an invented word. The model
treated them like any other turn. For example, it wrote the fact ``Caroline's new apartment door code is 28352,
and the building manager is called Thalwyndor.'' It copied the exact secret into a derived memory for
\LocoCanPropExact{} of the \LocoCanaries{} canaries, and the topic of the canary (``door code'', ``locker'', and
so on) into a descendant for \LocoCanPropTopic{}.

\paragraph{Schemes} We compare logical deletion with a soft-deleting HNSW index, per-object keys, epoch keys, and
\sys{} with the attributed and the sound rule (\cref{sec:impl}). The baselines erase the forgotten turn only,
which is their documented behaviour.

\subsection{E1--E2: do forgotten secrets survive?}\label{sec:e1}
The user asks to forget the \LocoCanaries{} canary turns. The attacker then obtains every file (database, index,
state) and the current secret state, and tries four channels: (i)~records it can read or decrypt that contain the
exact number or word; (ii)~the exact secret anywhere in the raw bytes of the files, which catches deleted rows
left in free pages; (iii)~the vectors it can read or decrypt: for each canary it ranks the true sentence against
99 decoys built from the same template with other numbers and words, and wins if the true one has the closest
vector, a strong attacker who knows the template, in the spirit of embedding inversion \cite{morris2023inversion};
(iv)~a readable descendant of the canary turn (ground truth: the model saw it) that mentions the canary's topic.

\begin{table*}[t]\centering\caption{Erasure of planted secrets after a real LLM wrote the memory (LoCoMo, 128 sessions, 35 canary turns forgotten). The attacker holds every file and the current secret state. Columns count canaries (out of 35) recovered through each channel.}\label{tab:canary}
\resizebox{\textwidth}{!}{%
\begin{tabular}{@{}lcccccc@{}}\toprule
Scheme & Exact secret in & Exact secret in & Identified from & Topic in a & Any & Objects\\
 & readable records & raw bytes & vectors (1 of 100) & descendant & channel & erased\\\midrule
Logical deletion + soft-deleted HNSW & 14 & 35 & 26 & 22 & \textbf{35} & 35\\
Per-object keys (no lineage) & 14 & 0 & 8 & 22 & \textbf{22} & 35\\
Epoch keys & 14 & 0 & 8 & 22 & \textbf{22} & 35\\
\sys{}, attributed rule & 0 & 0 & 0 & 3 & \textbf{3} & 56\\
\sys{}, sound rule & 0 & 0 & 0 & 0 & \textbf{0} & 526\\
\bottomrule\end{tabular}}\end{table*}

\Cref{tab:canary} gives the results. Logical deletion loses everything: the deleted turns stay in free SQLite
pages (\Canplainlogicalrecoveredfromrawbytes{} of \LocoCanaries{}) and their vectors come back after an undelete
of the soft-deleted HNSW points (\Canplainlogicalidentifiedfromvectors{} identified). Encryption with per-object
or epoch keys removes these storage residues, but not the model's copies: the summaries and facts written from the
canary turns are still readable, and leak the exact secret for \Canperobjectrecoveredbydecryption{} canaries, the
topic for \Canperobjecttopicleakindescendants{}, and enough for the vector attack in
\Canperobjectidentifiedfromvectors{} cases. In total \Canperobjectanychannel{} of the \LocoCanaries{} forgotten
canaries leak. \sys{} with the attributed rule never leaks an exact secret, but still leaks the topic of
\Canlineraattributedtopicleakindescendants{} canaries: the attribution function missed the canary turn when the
model folded it into a summary. \sys{} with the sound rule leaks nothing through any channel. It erases
\Canlinerasoundobjectslost{} objects to do so, against \Canlineraattributedobjectslost{} for the attributed rule:
this is the price we study next.

\subsection{E5: how far does a forget request reach?}\label{sec:e5}
\begin{table*}[t]\centering\caption{Lineage statistics and forgetting radius on the LoCoMo log. Descendants: derived objects erased by forgetting one turn. Radius: share of derived memories erased when a random fraction of turns is forgotten (mean of 3 seeds). $^\dagger$Structural what-if on the same log: the parent sets that the alternative pipeline would record.}\label{tab:lineage}
\begin{tabular}{@{}lccccccc@{}}\toprule
Lineage rule & Mean & \multicolumn{3}{c}{Descendants of one turn} & \multicolumn{3}{c}{Derived memories erased}\\
\cmidrule(lr){3-5}\cmidrule(l){6-8} & parents & mean & p95 & max & 1\% & 5\% & 10\%\\\midrule
Sound (as run, retrieves summaries) & 24.8 & 58.2 & 210 & 234 & 74.8\% & 99.0\% & 99.7\%\\
Attributed (top-3 context items) & 3.0 & 2.5 & 3 & 166 & 19.4\% & 38.0\% & 59.3\%\\
Flat derivation$^\dagger$ & -- & 27.4 & 89 & 221 & 59.7\% & 98.2\% & 99.7\%\\
Session-local$^\dagger$ & -- & 7.6 & 9 & 9 & 18.9\% & 65.7\% & 86.6\%\\
\bottomrule\end{tabular}\end{table*}
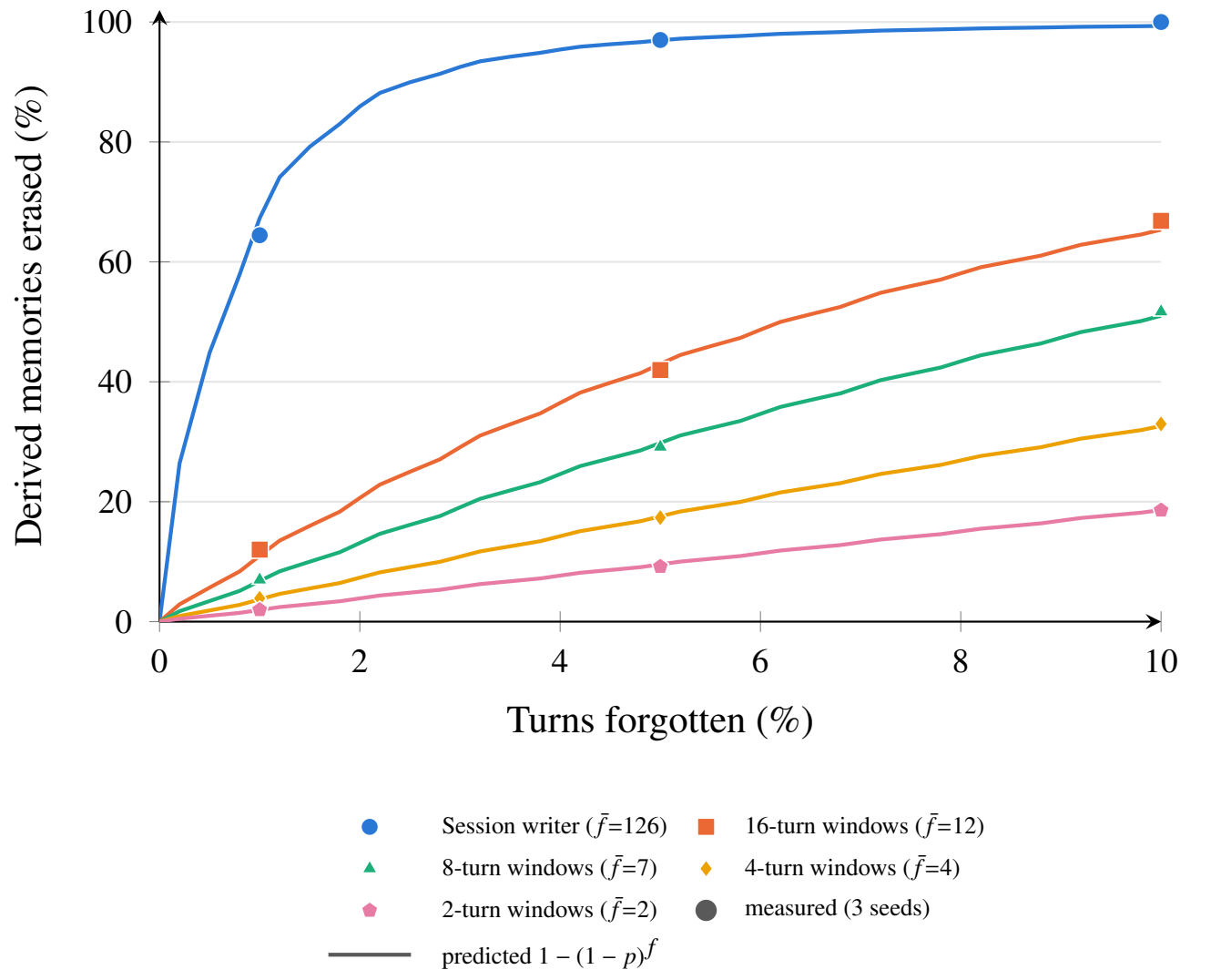
\begin{figure}[t]\centering
\begin{tikzpicture}\begin{axis}[width=\columnwidth, height=0.68\columnwidth, xmin=0, xmax=10, ymin=0, ymax=102,
  xlabel={Turns forgotten (\%)}, ylabel={Derived memories erased (\%)}, tick label style={font=\scriptsize},
  label style={font=\footnotesize}, ymajorgrids, grid style={black!10}, axis lines=left,
  legend style={font=\tiny, draw=none, fill=none, at={(0.5,-0.28)}, anchor=north, legend columns=2, column sep=4pt, row sep=-1pt}, legend cell align=left]
\definecolor{c0}{HTML}{2A78D6}
\addplot[c0, thick, forget plot] coordinates {(0.0,0.00) (0.2,26.40) (0.5,44.85) (0.8,57.91) (1.0,67.29) (1.2,74.14) (1.5,79.20) (1.8,83.01) (2.0,85.92) (2.2,88.17) (2.5,89.95) (2.8,91.36) (3.0,92.50) (3.2,93.44) (3.5,94.21) (3.8,94.86) (4.0,95.41) (4.2,95.88) (4.5,96.29) (4.8,96.64) (5.0,96.95) (5.2,97.22) (5.5,97.46) (5.8,97.67) (6.0,97.86) (6.2,98.03) (6.5,98.18) (6.8,98.32) (7.0,98.45) (7.2,98.57) (7.5,98.67) (7.8,98.77) (8.0,98.85) (8.2,98.93) (8.5,99.01) (8.8,99.08) (9.0,99.14) (9.2,99.20) (9.5,99.25) (9.8,99.30) (10.0,99.35)};
\addplot[only marks, mark=*, mark size=1.8pt, c0, mark options={draw=white, line width=0.3pt}] coordinates {(1,64.44) (5,97.00) (10,100.00)};
\addlegendentry{Session writer ($\bar f$=126)}
\definecolor{c1}{HTML}{EB6834}
\addplot[c1, thick, forget plot] coordinates {(0.0,0.00) (0.2,2.89) (0.5,5.68) (0.8,8.38) (1.0,10.99) (1.2,13.52) (1.5,15.97) (1.8,18.34) (2.0,20.63) (2.2,22.85) (2.5,24.99) (2.8,27.07) (3.0,29.08) (3.2,31.02) (3.5,32.90) (3.8,34.72) (4.0,36.48) (4.2,38.19) (4.5,39.84) (4.8,41.43) (5.0,42.98) (5.2,44.47) (5.5,45.92) (5.8,47.32) (6.0,48.68) (6.2,49.99) (6.5,51.26) (6.8,52.49) (7.0,53.69) (7.2,54.84) (7.5,55.96) (7.8,57.04) (8.0,58.09) (8.2,59.10) (8.5,60.08) (8.8,61.04) (9.0,61.96) (9.2,62.85) (9.5,63.72) (9.8,64.56) (10.0,65.37)};
\addplot[only marks, mark=square*, mark size=1.8pt, c1, mark options={draw=white, line width=0.3pt}] coordinates {(1,12.01) (5,41.96) (10,66.84)};
\addlegendentry{16-turn windows ($\bar f$=12)}
\definecolor{c2}{HTML}{1BAF7A}
\addplot[c2, thick, forget plot] coordinates {(0.0,0.00) (0.2,1.74) (0.5,3.44) (0.8,5.13) (1.0,6.78) (1.2,8.40) (1.5,10.00) (1.8,11.58) (2.0,13.12) (2.2,14.65) (2.5,16.14) (2.8,17.61) (3.0,19.06) (3.2,20.48) (3.5,21.88) (3.8,23.26) (4.0,24.61) (4.2,25.94) (4.5,27.25) (4.8,28.54) (5.0,29.80) (5.2,31.05) (5.5,32.27) (5.8,33.47) (6.0,34.65) (6.2,35.81) (6.5,36.96) (6.8,38.08) (7.0,39.18) (7.2,40.27) (7.5,41.33) (7.8,42.38) (8.0,43.41) (8.2,44.42) (8.5,45.41) (8.8,46.39) (9.0,47.35) (9.2,48.29) (9.5,49.22) (9.8,50.13) (10.0,51.02)};
\addplot[only marks, mark=triangle*, mark size=1.8pt, c2, mark options={draw=white, line width=0.3pt}] coordinates {(1,6.97) (5,29.13) (10,51.69)};
\addlegendentry{8-turn windows ($\bar f$=7)}
\definecolor{c3}{HTML}{EDA100}
\addplot[c3, thick, forget plot] coordinates {(0.0,0.00) (0.2,0.94) (0.5,1.87) (0.8,2.80) (1.0,3.72) (1.2,4.63) (1.5,5.54) (1.8,6.44) (2.0,7.34) (2.2,8.22) (2.5,9.10) (2.8,9.98) (3.0,10.84) (3.2,11.71) (3.5,12.56) (3.8,13.41) (4.0,14.25) (4.2,15.09) (4.5,15.92) (4.8,16.74) (5.0,17.56) (5.2,18.37) (5.5,19.17) (5.8,19.97) (6.0,20.76) (6.2,21.55) (6.5,22.33) (6.8,23.10) (7.0,23.87) (7.2,24.63) (7.5,25.39) (7.8,26.14) (8.0,26.89) (8.2,27.63) (8.5,28.36) (8.8,29.09) (9.0,29.81) (9.2,30.52) (9.5,31.23) (9.8,31.94) (10.0,32.64)};
\addplot[only marks, mark=diamond*, mark size=1.8pt, c3, mark options={draw=white, line width=0.3pt}] coordinates {(1,3.89) (5,17.35) (10,32.97)};
\addlegendentry{4-turn windows ($\bar f$=4)}
\definecolor{c4}{HTML}{E87BA4}
\addplot[c4, thick, forget plot] coordinates {(0.0,0.00) (0.2,0.49) (0.5,0.98) (0.8,1.46) (1.0,1.95) (1.2,2.43) (1.5,2.91) (1.8,3.40) (2.0,3.88) (2.2,4.36) (2.5,4.83) (2.8,5.31) (3.0,5.79) (3.2,6.26) (3.5,6.73) (3.8,7.20) (4.0,7.68) (4.2,8.15) (4.5,8.61) (4.8,9.08) (5.0,9.55) (5.2,10.01) (5.5,10.48) (5.8,10.94) (6.0,11.40) (6.2,11.86) (6.5,12.32) (6.8,12.77) (7.0,13.23) (7.2,13.69) (7.5,14.14) (7.8,14.59) (8.0,15.05) (8.2,15.50) (8.5,15.94) (8.8,16.39) (9.0,16.84) (9.2,17.29) (9.5,17.73) (9.8,18.17) (10.0,18.61)};
\addplot[only marks, mark=pentagon*, mark size=1.8pt, c4, mark options={draw=white, line width=0.3pt}] coordinates {(1,1.98) (5,9.19) (10,18.58)};
\addlegendentry{2-turn windows ($\bar f$=2)}
\addlegendimage{only marks, mark=*, black!65}\addlegendentry{measured (3 seeds)}
\addlegendimage{black!65, thick}\addlegendentry{predicted $1-(1-p)^f$}
\end{axis}\end{tikzpicture}
\caption{Share of derived memories erased by the sound rule when a random fraction of turns is forgotten, for writers
with different context sizes (LoCoMo conversations 0--1, same LLM). Markers: measured (mean of 3 seeds). Lines: the
fan-in law $1-(1-p)^f$ averaged over the measured fan-in distribution, with no fitted parameter.}\label{fig:radius}
\end{figure}

\Cref{tab:lineage} describes the lineage graph that the model's prompts induce. Under the sound rule a derived
memory has \Linsoundmeanparents{} parents on average and, through retrieved summaries, \Linsoundmeansourceancestors{}
source ancestors. Forgetting one turn erases \Linsoundsingleforgetdescendantsmean{} derived memories on average.
When a random 1\% of the turns is forgotten, \RadsoundZeroZeroOne{} of all derived memories disappear; at 5\%,
almost all of them. Two structural what-ifs on the same log show where this comes from. Flat derivation (the
writer is shown the turns of the retrieved sessions instead of their summaries) removes chains, but each memory
still has about a hundred source parents, and the radius at 1\% only drops to \RadflatZeroZeroOne{}. What matters
is the number of source turns in the writer's context, its \emph{fan-in}~$f$: a memory survives a random forget
rate $p$ with probability about $(1-p)^{f}$ (\cref{fig:radius}). A typical
session writer has $f$ in the hundreds once retrieval is counted.

The remedy is to keep the writer's context small. We ran the same model a second time on conversations 0 and 1,
now extracting up to three facts from each window of four consecutive turns, with no retrieval of earlier memories
(\RrwindowFourwriterconvZeroOnellmcalls{} short calls instead of 38 long ones). The sound rule is unchanged, but the
mean fan-in drops from \RrsessionwriterconvZeroOnesoundmeansourceancestors{} to
\RrwindowFourwriterconvZeroOnesoundmeansourceancestors{} source turns, and the forgetting radius at 1\% falls from
\RrsessionwriterconvZeroOnesounderasedZeroZeroOne{} to \RrwindowFourwriterconvZeroOnesounderasedZeroZeroOne{}
(\RrwindowFourwriterconvZeroOnesounderasedZeroOne{} at 10\%, against \RrsessionwriterconvZeroOnesounderasedZeroOne{}).
The fan-in law predicts these values closely: \RrwindowFourwriterconvZeroOnesoundpredictedZeroZeroOne{} and
\RrsessionwriterconvZeroOnesoundpredictedZeroZeroOne{} at 1\%. Security does not change: on the
\WinCanaries{} canaries of these conversations, the windowed writer copied the exact secret into a fact
\WinCanPropExact{} times; logical deletion leaks \WinCanplainlogicalanychannel{} of them, per-object and epoch keys
\WinCanperobjectanychannel{}, and \sys{} with the sound rule none, while erasing only
\WinCanlinerasoundobjectslost{} objects.

\subsection{E3: what does the guarantee cost in answer quality?}\label{sec:e3}
We forget a random 1\%, 5\% or 10\% of the turns (nested sets, three seeds) and score only the questions whose
evidence turns all survive: the others are \emph{meant} to become unanswerable. We use two views of the store:
\emph{facts-only}, where the agent retrieves derived memories only (as mem0 does), and \emph{hybrid}, where the
encrypted raw turns are also retrievable.

\begin{table*}[t]\centering\caption{Retrieval utility on LoCoMo after forgetting: share of retained questions for which a top-10 hit comes from an evidence session (mean of 3 seeds). Facts-only: the store keeps derived memories only (mem0-like); hybrid: raw turns are kept too.}\label{tab:retrieval}
\setlength{\tabcolsep}{3pt}\begin{tabular}{@{}lcccccc@{}}\toprule
& \multicolumn{3}{c}{Facts-only} & \multicolumn{3}{c}{Hybrid}\\\cmidrule(lr){2-4}\cmidrule(l){5-7}
Condition & 1\% & 5\% & 10\% & 1\% & 5\% & 10\%\\\midrule
No forgetting & \multicolumn{3}{c}{63.2\%} & \multicolumn{3}{c}{79.5\%}\\
No lineage (logical, per-object, epoch) & 63.1\% & 63.0\% & 62.9\% & 79.4\% & 79.3\% & 79.5\%\\
\sys{}, attributed & 61.6\% & 59.0\% & 54.1\% & 79.4\% & 78.8\% & 78.5\%\\
\sys{}, sound & 18.9\% & 1.8\% & 0.4\% & 70.2\% & 67.6\% & 68.3\%\\
\bottomrule\end{tabular}\end{table*}

\paragraph{Retrieval} \Cref{tab:retrieval} reports how often a top-10 hit comes from an evidence session, over
\LocoQuestions{} questions (category 5, adversarial, excluded). Schemes without lineage lose nothing, because they
erase almost nothing. The sound rule makes the facts-only store collapse, from \RetpZeroZerononefactsonly{} to
\RetpZeroZeroOnesoundfactsonly{} at 1\%: this is the radius of \cref{sec:e5}. The hybrid store degrades much less
(\RetpZeroZerononehybrid{} to \RetpZeroZeroOnesoundhybrid{} at 1\%, \RetpZeroOnesoundhybrid{} at 10\%),
because surviving raw turns still carry most of the evidence. The attributed rule keeps utility close to the
baseline, but, as \cref{tab:canary} showed, at the price of leaks.

\begin{table*}[t]\centering\caption{Answer quality from the surviving memory (token F1, \%), share of ``unknown'' answers (\%) and number of sessions re-summarised. Same LLM for writing, re-deriving and answering; only questions whose evidence survives are scored.}\label{tab:qa}
\setlength{\tabcolsep}{2.5pt}\begin{tabular}{@{}lcccccc@{}}\toprule
& \multicolumn{3}{c}{LoCoMo (conv.~0--1)} & \multicolumn{3}{c}{LongMemEval-S subset}\\\cmidrule(lr){2-4}\cmidrule(l){5-7}
Condition & F1 & unk. & re-der. & F1 & unk. & re-der.\\\midrule
No forgetting & 12.9 & 0 & -- & 20.0 & 0 & --\\
No lineage, 1\% & 12.9 & 0 & -- & 20.0 & 0 & --\\
\sys{} attributed, 1\% & 12.9 & 0 & -- & 20.0 & 0 & --\\
\sys{} sound, 1\% & 10.5 & 0 & -- & 18.8 & 0 & --\\
\sys{} sound + re-derivation, 1\% & 17.2 & 0 & 26 & 20.1 & 0 & 4\\
No lineage, 10\% & 12.9 & 0 & -- & 20.0 & 0 & --\\
\sys{} attributed, 10\% & 11.6 & 0 & -- & 23.8 & 0 & --\\
\sys{} sound, 10\% & 7.1 & 0 & -- & 12.6 & 0 & --\\
\sys{} sound + re-derivation, 10\% & 11.7 & 0 & 38 & 24.3 & 0 & 53\\
\bottomrule\end{tabular}\end{table*}

\paragraph{Answers} \Cref{tab:qa} closes the loop with the same LLM answering \QALoN{} sampled questions from
the facts-only store of conversations 0 and 1. The small model sets a low ceiling (token F1
\QALononeatZeropZeroFone{} without forgetting, bootstrap standard deviation \QALononeatZeropZeroSD{}) and a floor
of \QALosoundatZeropOneFone{} when the store is empty, which is the case for the sound rule at 10\%. At 1\% the
sound rule keeps \QALosoundatZeropZeroOneAlive{} of 300 derived memories and F1 drops to
\QALosoundatZeropZeroOneFone{}. \emph{Re-derivation removes the loss}: after re-summarising the
\QALosoundrdatZeropZeroOneCalls{} affected sessions (\QALosoundrdatZeropZeroOneMin{}~minutes on CPU), F1 is
\QALosoundrdatZeropZeroOneFone{} at 1\% and \QALosoundrdatZeropOneFone{} at 10\%
(\QALosoundrdatZeropOneCalls{} sessions, \QALosoundrdatZeropOneMin{}~minutes): above the no-forgetting level at 1\%,
and within one standard deviation of it at 10\%.
Re-derived memories are written from fewer turns and appear slightly easier to retrieve; we do not read more into
this difference, which is within two standard deviations. The guarantee is kept: re-derived memories have only
surviving parents.

\paragraph{Small-context writing} The windowed writer of \cref{sec:e5} changes the picture. On the same 60
questions it answers \emph{better} than the session writer even without forgetting (F1 \QAWinnoneatZeropZeroFone{}
against \QALononeatZeropZeroFone{}): short excerpts give the small model less to summarise away. Under the sound
rule, forgetting 1\% of the turns leaves F1 at \QAWinsoundatZeropZeroOneFone{}, and 10\% at
\QAWinsoundatZeropOneFone{} (\QAWinsoundatZeropOneAlive{} of \QAWinsoundatZeropOneOrig{} facts survive).
Re-derivation is cheap because only the windows that lost a turn are rewritten: \QAWinsoundrdatZeropZeroOneCalls{}
short calls at 1\% and \QAWinsoundrdatZeropOneCalls{} at 10\% (\QAWinsoundrdatZeropOneMin{}~minutes), giving F1
\QAWinsoundrdatZeropZeroOneFone{} and \QAWinsoundrdatZeropOneFone{}. In short, with a small-context writer the
sound rule costs almost nothing in answer quality.

\paragraph{LongMemEval} We repeated phases A and B on a seeded subset of LongMemEval-S \cite{wu2025longmemeval}:
six questions of each of four types (multi-session, temporal reasoning, knowledge update, single-session user),
each with its own history made of its evidence sessions and one distractor, user turns only
(\LmeSessions{} sessions, \LmeSources{} turns, \LmeDerived{} derived memories). These histories are short, so the
sound-rule radius is smaller than on LoCoMo (\LmeRadsoundZeroZeroOne{} of derived memories at 1\%,
\LmeRadsoundZeroOne{} at 10\%) and retrieval of evidence sessions saturates. The answer-quality trend is the same
(right half of \cref{tab:qa}): the sound rule costs little at 1\% and much at 10\%, and re-derivation recovers it.
With \QALmeN{} questions the standard deviation is about \QALmenoneatZeropZeroSD{} points, so we read only the
direction of these differences.

\paragraph{Size of the writer context} \Cref{tab:ablation} and \cref{fig:radius} vary the window from 2 to 16
turns on conversations 0 and 1. The fan-in law holds at every size, with no fitted parameter. The radius at 1\%
grows with the window, from \AbwindowTwowriterconvZeroOneerasedZeroZeroOne{} (2 turns) to
\AbwindowOneSixwriterconvZeroOneerasedZeroZeroOne{} (16 turns), against \AbsessionwriterconvZeroOneerasedZeroZeroOne{}
for the session writer. Answer quality does not fall when the window shrinks: F1 without forgetting is
\AbwindowTwowriterconvZeroOneFonenoneZeroZero{} with 2-turn windows and \AbwindowFourwriterconvZeroOneFonenoneZeroZero{}
with 4, against \AbwindowEightwriterconvZeroOneFonenoneZeroZero{} with 8, \AbwindowOneSixwriterconvZeroOneFonenoneZeroZero{}
with 16 and \AbsessionwriterconvZeroOneFonenoneZeroZero{} for whole sessions. In every configuration
\sys{} leaks no canary, while per-object keys leak most of them. Note the tension for schemes without lineage:
small-context writers copy secrets verbatim more often (per-object keys leak
\AbwindowTwowriterconvZeroOneCanperobject{} of \AbwindowTwowriterconvZeroOneCanaries{} canaries with 2-turn windows,
\AbsessionwriterconvZeroOneCanperobject{} with the session writer), so the configuration that is best for utility is
also the one where lineage matters most.

\begin{table*}[t]\centering\caption{Writer context size and model family (LoCoMo conversations 0--1, sound rule). Fan-in: mean number of source turns behind a derived memory. Erased: share of derived memories erased when a random 1\% or 10\% of turns is forgotten (3 seeds); predicted: fan-in law. F1: answers from the facts-only store (60 questions); rd: after re-derivation. Leaks: forgotten canaries recovered by any channel with per-object keys \textbar{} with \sys{}.}\label{tab:ablation}
\setlength{\tabcolsep}{3.5pt}\resizebox{\textwidth}{!}{%
\begin{tabular}{@{}lrrrrrrrrrrc@{}}\toprule
Writer & LLM & Derived & Fan-in & \multicolumn{2}{c}{Erased at 1\%} & Erased & \multicolumn{4}{c}{F1} & Leaks\\
\cmidrule(lr){5-6}\cmidrule(lr){8-11} & calls & memories & & measured & predicted & at 10\% & none & 1\% & 1\%+rd & 10\% & \\\midrule
Session writer (Qwen2.5-1.5B) & 38 & 300 & 126.2 & 64.4\% & 67.3\% & 100.0\% & 12.9 & 10.5 & 17.2 & 7.1 & 7/11 \,|\, 0/11\\
16-turn windows (Qwen2.5-1.5B) & 70 & 197 & 11.7 & 12.0\% & 11.0\% & 66.8\% & 18.2 & 16.4 & -- & 11.3 & 6/11 \,|\, 0/11\\
8-turn windows (Qwen2.5-1.5B) & 116 & 325 & 7.0 & 7.0\% & 6.8\% & 51.7\% & 20.7 & 21.6 & -- & 17.8 & 8/11 \,|\, 0/11\\
4-turn windows (Qwen2.5-1.5B) & 214 & 557 & 3.8 & 3.9\% & 3.7\% & 33.0\% & 24.1 & 24.1 & 24.2 & 21.3 & 10/11 \,|\, 0/11\\
2-turn windows (Qwen2.5-1.5B) & 409 & 958 & 2.0 & 2.0\% & 1.9\% & 18.6\% & 23.5 & 23.3 & -- & 18.6 & 11/11 \,|\, 0/11\\
Session writer (Llama-3.2-3B) & 38 & 342 & 141.9 & 63.2\% & 70.7\% & 100.0\% & 28.5 & 16.1 & 26.9 & 1.7 & 10/11 \,|\, 0/11\\
\bottomrule\end{tabular}}\end{table*}

\paragraph{A second model family} To check that these effects do not depend on one model, we replayed conversations 0
and 1 with Llama-3.2-3B-Instruct as the session writer (same prompts, same canaries). This model copies secrets
verbatim far more often: the exact secret reached a derived memory for \AbllamaThreebsessionwriterconvZeroOneCanProp{}
of \AbllamaThreebsessionwriterconvZeroOneCanaries{} canaries. Per-object and epoch keys therefore leak
\AbllamaThreebsessionwriterconvZeroOneCanperobject{} of them, logical deletion
\AbllamaThreebsessionwriterconvZeroOneCanplainlogical{}, the attributed rule
\AbllamaThreebsessionwriterconvZeroOneCanlineraattributed{}, and the sound rule none. Its fan-in
(\AbllamaThreebsessionwriterconvZeroOnemeansourceancestors{} source turns) gives a radius of
\AbllamaThreebsessionwriterconvZeroOneerasedZeroZeroOne{} at 1\%, close to the law's
\AbllamaThreebsessionwriterconvZeroOnepredictedZeroZeroOne{}. The stronger model also makes the utility trade-off
clearer: F1 is \AbllamaThreebsessionwriterconvZeroOneFonenoneZeroZero{} without forgetting,
\AbllamaThreebsessionwriterconvZeroOneFonesoundZeroZeroOne{} under the sound rule at 1\%, and
\AbllamaThreebsessionwriterconvZeroOneFonesoundrdZeroZeroOne{} after re-derivation.

\subsection{A larger writer on all conversations}\label{sec:e7b}
\begin{table*}[t]\centering\caption{Extended campaign: Qwen2.5-7B-Instruct writes, re-derives and answers (vLLM, one GPU). LoCoMo: all ten conversations; LongMemEval-S: 100 histories. Fan-in: mean number of source turns behind a derived memory (sound rule). Leaks: forgotten canaries recovered by any channel (\cref{sec:e1}). Erased: share of derived memories erased by the sound rule when a random 1\% or 10\% of turns is forgotten (3 seeds); predicted: fan-in law. F1: token F1 (\%) from the facts-only store over every eligible question; rd: after re-derivation.}\label{tab:qwen7b}
\setlength{\tabcolsep}{2.6pt}\resizebox{\textwidth}{!}{%
\begin{tabular}{@{}lrrcccccccrrrrr@{}}\toprule
Writer & Derived & Fan-in & \multicolumn{4}{c}{Leaks} & \multicolumn{2}{c}{Erased at 1\%} & Erased & \multicolumn{5}{c}{F1}\\
\cmidrule(lr){4-7}\cmidrule(lr){8-9}\cmidrule(l){11-15} & memories & & logical & obj./ep. & attr. & sound & meas. & pred. & at 10\% & none & 1\% & 1\%+rd & 10\% & 10\%+rd\\\midrule
LoCoMo, session writer & 2{,}229 & 181.5 & 70 & 67 & 13 & \textbf{0} & 73.0\% & 77.2\% & 99.6\% & 27.8 & 13.8 & 27.9 & 2.0 & 27.3\\
LoCoMo, 4-turn windows & 4{,}172 & 3.8 & 70 & 69 & 0 & \textbf{0} & 3.8\% & 3.8\% & 32.7\% & 31.8 & 30.7 & 31.8 & 26.6 & 31.9\\
LongMemEval-S, session writer & 2{,}247 & 11.1 & -- & -- & -- & -- & 10.4\% & -- & 65.5\% & 27.8 & 26.9 & 28.7 & 12.4 & 29.7\\
\bottomrule\end{tabular}}\end{table*}
The runs above use small models on CPU. We replayed the whole pipeline with Qwen2.5-7B-Instruct on one GPU, on all
ten LoCoMo conversations (\QSevenSesSessions{} sessions, \QSevenSesSources{} turns, \QSevenSesCanaries{} canaries
from the same templates and seed), with both the session writer and four-turn windows, and on 100 LongMemEval-S
histories (25 questions of each of the four types). Questions are no longer sampled: we score every question whose
evidence survives. \Cref{tab:qwen7b} summarises E1--E5 for this writer.

\paragraph{Leaks} The larger model copies the exact secret into a derived memory for \QSevenSesCanProp{} of the
\QSevenSesCanaries{} canaries (\QSevenWinCanProp{} with four-turn windows). Logical deletion leaks
\QSevenSesCanplainlogicalanychannel{} of them through some channel, and per-object and epoch keys
\QSevenSesCanperobjectanychannel{}, entirely through the model's copies (the exact secret is readable in a derived
memory for \QSevenSesCanperobjectrecoveredbydecryption{} of them). The attributed rule leaks the topic of
\QSevenSesCanlineraattributedanychannel{} canaries with the session writer and none with windows. The sound rule leaks
none in either setting.

\paragraph{Radius} The session writer has a mean fan-in of \QSevenSesFanin{} source turns, and forgetting 1\% of the
turns erases \QSevenSesRadsoundZeroZeroOne{} of the derived memories, against \QSevenSesPredZeroZeroOne{} predicted by
the fan-in law. Four-turn windows bring the fan-in down to \QSevenWinFanin{} and the radius to
\QSevenWinRadsoundZeroZeroOne{} (predicted \QSevenWinPredZeroZeroOne{}), and to \QSevenWinRadsoundZeroOne{} at 10\%.
The law thus holds, with no fitted parameter, for another model size, another inference engine and five times more
history.

\paragraph{Answers} Over \QSevenSesQAN{} LoCoMo questions (bootstrap standard deviation about
\QSevenSesQAnoneatZeropZeroSD{} points), the session writer reaches F1 \QSevenSesQAnoneatZeropZeroFone{} without
forgetting. The sound rule drops it to \QSevenSesQAsoundatZeropZeroOneFone{} at 1\%, where
\QSevenSesQAsoundatZeropZeroOneAlive{} of \QSevenSesDerived{} derived memories survive, and to
\QSevenSesQAsoundatZeropOneFone{} at 10\%, where none does. Re-derivation restores it: F1
\QSevenSesQAsoundrdatZeropZeroOneFone{} at 1\% (\QSevenSesQAsoundrdatZeropZeroOneCalls{} sessions,
\QSevenSesQAsoundrdatZeropZeroOneMin{}~minutes) and \QSevenSesQAsoundrdatZeropOneFone{} at 10\%
(\QSevenSesQAsoundrdatZeropOneCalls{} sessions, \QSevenSesQAsoundrdatZeropOneMin{}~minutes). As with the small model,
the windowed writer answers better even without forgetting (\QSevenWinQAnoneatZeropZeroFone{}), keeps
\QSevenWinQAsoundatZeropZeroOneFone{} at 1\% and \QSevenWinQAsoundatZeropOneFone{} at 10\% under the sound rule, and
returns to \QSevenWinQAsoundrdatZeropZeroOneFone{} and \QSevenWinQAsoundrdatZeropOneFone{} after re-deriving
\QSevenWinQAsoundrdatZeropZeroOneCalls{} and \QSevenWinQAsoundrdatZeropOneCalls{} short windows. On LongMemEval-S
(\QSevenLmeQAN{} eligible questions, standard deviation about \QSevenLmeQAnoneatZeropZeroSD{} points), the sound rule
costs little at 1\% (\QSevenLmeQAsoundatZeropZeroOneFone{} against \QSevenLmeQAnoneatZeropZeroFone{}) and much at 10\%
(\QSevenLmeQAsoundatZeropOneFone{}), and re-derivation brings F1 to \QSevenLmeQAsoundrdatZeropZeroOneFone{} and
\QSevenLmeQAsoundrdatZeropOneFone{}.

A larger writer answers better but changes none of our conclusions: what leaks after deletion is the model's own
copies, the sound rule leaks nothing, the radius follows the fan-in law, and small windows with re-derivation make the
guarantee almost free in answer quality.

\subsection{Case study: mem0}\label{sec:mem0}
We ran mem0 2.2.1 end to end, with the same local model as its fact extractor, on six user messages, two of which
contain secrets (a door code with the manager's name, and a vet confirmation number). The extraction prompt of
mem0 alone is longer than 8{,}000 tokens, so this experiment uses a 16k-token context.

\paragraph{Vanilla mem0} With its default on-disk stores (Qdrant and a SQLite history), mem0 wrote \MemVanBefore{}
memories. We served the erasure request the way mem0 offers it: search the memories about each forgotten topic
and call \texttt{delete} on every match. This removed \MemVanDeleted{} memories, and no live memory mentions a
secret any more. Yet all \MemSecrets{} secrets are still in the files, in \MemVanFiles{}: the history database keeps
the text of deleted memories by design, and the SQLite file of the vector store still contains them.

\paragraph{mem0 with \sys{}} With the adapter of \cref{sec:construction}, forgetting the two messages erased
\MemLinClosure{} objects. An attacker holding every file and the current state finds no secret, neither in
decryptable records (\MemLinSecretsDec{}) nor in raw bytes (\MemLinSecretsFiles{}). The run also shows the radius
at work in a real framework: because mem0 shows its extractor the last ten messages, every memory written after a
forgotten message has it as a parent, and only \MemLinAfter{} of the \MemLinBefore{} memories survives. With mem0's default
context, honouring erasure soundly means re-deriving; a smaller extraction window, as in \cref{sec:e5}, makes this
cheap.

\paragraph{Longer run, larger extractor} We repeated the case study with Qwen2.5-7B-Instruct as mem0's extractor and
\QSevenMemMessages{} user messages. Vanilla mem0 deleted \QSevenMemVanDeleted{} memories on request, and
\QSevenMemVanSecretsFiles{} secrets were still in \QSevenMemVanFiles{} afterwards. With \sys{}, forgetting erased
\QSevenMemLinClosure{} objects, only \QSevenMemLinAfter{} of the \QSevenMemLinBefore{} memories survived, and the
attacker found no secret, neither in decryptable records (\QSevenMemLinSecretsDec{}) nor in raw bytes
(\QSevenMemLinSecretsFiles{}).

\subsection{E4: system costs}\label{sec:e4}
\begin{table*}[t]\centering\caption{System costs of \sys{} (Python, one process on the 2-core Intel Xeon at 2.1~GHz). Write includes the vector index; search is a warm top-10 scan over all shards; forget includes the closure, the puncture, the atomic state write and the shard rebuilds (median of 100 random source turns); state is the PPRF key after these forgets.}\label{tab:scale}
\begin{tabular}{@{}lrrrrrrrrr@{}}\toprule
Regime & Objects & Write & Read & Search & Forget & Descendants & State & Rotate & DB\\
 & & (ms) & (ms) & (ms) & (ms) & (median) & (KiB) & (s) & (MiB)\\\midrule
Chained & 10{,}013 & 0.13 & 0.16 & 2 & 11 & 1400 & 95.7 & 0.3 & 18\\
Flat & 10{,}013 & 0.13 & 0.13 & 2 & 7 & 9 & 95.8 & 0.4 & 18\\
Chained & 100{,}006 & 0.16 & 1.36 & 30 & 60 & 13023 & 95.2 & 2.9 & 177\\
Flat & 100{,}006 & 0.19 & 0.17 & 28 & 7 & 9 & 95.9 & 5.1 & 178\\
Flat & 1{,}000{,}029 & 0.28 & 0.16 & 326 & 9 & 9 & 95.4 & 59.9 & 1775\\
\bottomrule\end{tabular}\end{table*}
\Cref{tab:scale} measures the library on synthetic workloads shaped like our LLM log (sessions of 22 sources and 9
derived objects with 384-dimensional vectors), up to a million objects, in plain Python. Writing an object, its
key derivation and its vector costs \ScflatOneZeroZeroZeroZeroZeroZerowritemsperobject{}~ms at $10^6$ objects, and
reading it with a cold key cache \ScflatOneZeroZeroZeroZeroZeroZerogetmscold{}~ms. A forget request, which includes
the closure, the puncture, the atomic state write and the shard rebuilds, takes
\ScflatOneZeroZeroZeroZeroZeroZeroforgetmsmedian{}~ms (median) at $10^6$ objects. After 100 forgets the PPRF state
is \ScflatOneZeroZeroZeroZeroZeroZeropprfstateKiBafterforgets{}~KiB, and compaction of a million objects takes
\ScflatOneZeroZeroZeroZeroZeroZerorotates{}~s and brings the state back to a single node. The warm search scan over
all shards (\ScflatOneZeroZeroZeroZeroZeroZerosearchwarmms{}~ms at $10^6$) is the main cost and would use an
approximate index per shard in production.

The two regimes confirm \cref{sec:e5} at scale. With flat lineage (sources only as parents) the closure of a
forgotten turn stays at \ScflatOneZeroZeroZeroZeroZeroZeroclosuredescendantsmedian{} objects from $10^4$ to $10^6$
objects. With chained lineage (the writer retrieves earlier derived memories) it grows with the history, to a
median of \ScchainedOneZeroZeroZeroZeroZeroclosuredescendantsmedian{} objects at $10^5$, and a forget takes
\ScchainedOneZeroZeroZeroZeroZeroforgetmsmedian{}~ms. In our real LLM run, \sys{} added about
\CanlinerasoundWriteMs{}~ms per written object and \CanlinerasoundForgetMs{}~ms per forget request, negligible next
to the \LocoLLMHours{}~hours the model spent writing the memory.
\section{Discussion}\label{sec:discussion}

\paragraph{Choice of lineage rule}
The sound rule is the only one with a guarantee, and it costs nothing to implement: the hook sees the prompt.
Its price is the forgetting radius, and our measurements give a simple design rule for memory pipelines that
must honour erasure requests: \emph{keep the writer's context small}. A memory written from $f$ source turns
survives a forget rate $p$ with probability about $(1-p)^f$; session summaries that also read older summaries
have $f$ in the hundreds, four-turn windows have $f\approx4$. In our runs the small-context writer was also the
more useful one, and re-deriving after forgetting was cheap. The attributed rule is useful
as a utility knob, but a deployment that uses it should not claim erasure of derived data.

\paragraph{Reading Article~17}
The GDPR gives data subjects a right to erasure ``without undue delay'' \cite{gdpr2016}. For an agent memory,
the hard part is not the deletion of the message the user points at, but of what the system inferred from it.
\sys{} turns that into a property of the storage layer: once the request is processed, the memory operator can
show that no key exists, anywhere, that decrypts the forgotten message or its derivatives, including in backups.
This is a stronger statement than a deletion receipt \cite{singhal2026forgetagent}, and it holds against the
operator itself after the fact. It does not cover the model weights, which is the domain of machine unlearning
\cite{bourtoule2021unlearning}, nor other sources that repeat the same fact.

\paragraph{Deployment}
In a hosted memory service the state $S$ is small (about one kilobyte per forget request before compaction, \cref{sec:e4}) and can
live in a hardware security module or a TPM-sealed file, while the encrypted store stays in ordinary databases and
backups. Because forgetting is a key operation, it reaches backups that the operator cannot rewrite, which is the
classic argument for cryptographic erasure \cite{reardon2013sok}. Multi-tenant services keep one state per user.

\paragraph{Limitations}
Our LLM experiments use open models of at most 7B parameters: Qwen2.5-1.5B on five LoCoMo conversations (two for
the main question-answering study, with 60 sampled questions) and a 24-question subset of LongMemEval-S,
Llama-3.2-3B on two conversations, and Qwen2.5-7B on all ten conversations and 100 LongMemEval-S histories. Frontier
models may copy secrets verbatim more or less often, which changes how much the baselines leak but not the guarantee
of \sys{}. Exact answers also depend on the inference engine and hardware: rerunning the same quantised model on
another CPU changed a few answers, so we compare conditions only within a table computed on one machine. The lineage graph leaks its shape. Lineage capture assumes that
the hook sees the whole prompt; tools that read memory outside the framework would bypass it. Finally, the
attributed rule we test is deliberately simple; better attribution would raise utility but would still not give a
guarantee.
\section{Conclusion}\label{sec:conclusion}
Agent memories are mostly made of what the agent wrote itself. Forgetting a user's input therefore means forgetting
its lineage. We defined this property against an adversary that later steals the storage and the current keys,
built \sys{} from a puncturable PRF and a key graph that follows the lineage, proved it secure, and showed on a real
LLM pipeline that the usual deletion methods leave the forgotten information readable while \sys{} does not. The
price is a larger forgetting radius, which follows a simple fan-in law and which small-context writing and re-derivation bring down sharply. Two questions remain
open: hiding the shape of the lineage graph, and connecting storage-level erasure with the unlearning of model
weights.

\section*{Artifact availability}
The implementation of \sys{}, the baselines, the canary generator, the logs of every LLM call (prompts, outputs and
model file digests) and the scripts that produce every number and table of this paper are publicly available at
\url{https://github.com/yulliwasameur/linera} and archived on Zenodo \cite{ameur2026lineracode} (doi:\href{https://doi.org/10.5281/zenodo.24166476}{10.5281/zenodo.24166476}).
The code is released under the MIT licence. The logs contain text derived from LoCoMo and LongMemEval and are
released under the terms of those datasets; the datasets themselves are not redistributed, and the scripts download
them from their original sources. All runs use seed 20261109.

\section*{Data availability}
The code, the logs of every LLM call and the scripts that produce every number of this paper are deposited on
Zenodo \cite{ameur2026lineracode} (\href{https://doi.org/10.5281/zenodo.24166476}{doi:10.5281/zenodo.24166476})
and mirrored at \url{https://github.com/yulliwasameur/linera}. LoCoMo and LongMemEval are public datasets and are not
redistributed.

\section*{CRediT authorship contribution statement}
\textbf{Yulliwas Ameur:} Conceptualization, Methodology, Software, Formal analysis, Investigation, Data curation,
Visualization, Writing -- original draft, Writing -- review \& editing.
\textbf{Billal Mokhtari:} Investigation, Software, Validation, Data curation, Writing -- review \& editing.
\textbf{Samia Bouzefrane:} Supervision, Writing -- review \& editing.

\section*{Declaration of competing interest}
The authors declare that they have no known competing financial interests or personal relationships that could
have appeared to influence the work reported in this paper.

\section*{Funding}
This work was supported by a French government grant managed by the Agence Nationale de la Recherche under the
``Investissements d'avenir'' program (reference ``ANR-21-ESRE-0051''), through the MesoNET computing resources
used for the GPU experiments. The funding source had no involvement in the study design, in the collection, analysis
and interpretation of data, in the writing of the report, or in the decision to submit the article for publication.

\section*{Acknowledgements}
Centre de Calcul Intensif d'Aix-Marseille is acknowledged for granting access to its high performance computing
resources.

\section*{Declaration of generative AI and AI-assisted technologies in the manuscript preparation process}
During the preparation of this work the authors used Claude (Anthropic) to assist with writing code and experiment
scripts, and with drafting, editing and formatting parts of the text. The research questions, the reading and
analysis of the literature, and the scientific choices and conclusions are the authors' own. After using this tool,
the authors reviewed and edited the content as needed and take full responsibility for the content of the published
article.

\bibliographystyle{elsarticle-num}
\bibliography{refs}

\appendix
\section{Proof of Theorem~\ref{thm:main}}\label{app:proof}

We write $\Pr[H_i]$ for the probability that $b'=b$ in game $H_i$, so that
$\mathrm{Adv}^{\mathrm{lcfs}}_{\sys}(\mathcal{A})=2\,|\Pr[H_0]-\tfrac12|$. All oracles are lazily sampled tables
shared by the challenger and $\mathcal{A}$; the challenger's own queries are not visible to $\mathcal{A}$.

\paragraph{$H_0$} The real game $\mathbf{G}^{\mathrm{lcfs}}_{\sys}(\mathcal{A})$. A \textsc{Rotate} call replaces
the whole PPRF key by a fresh one; since a discarded key is never given to $\mathcal{A}$, we treat it below as a key
in which every point is punctured.

\paragraph{$H_1$: punctured leaves become uniform}
Call a GGM node \emph{hidden} if it is a prefix of a punctured tag (root and leaf included) and does not belong to
the key returned by \textsc{Corrupt}. There are at most $|\Phi|(n+1)$ hidden nodes per key. A hidden node is either
the root, which is uniform, or the output $G_b(s)$ of a hidden parent $s$. Revealed nodes are children of hidden nodes
(the siblings kept by \textsf{Puncture}) or descend from revealed nodes. In $H_1$ the challenger answers every
evaluation that passes through a hidden node with fresh uniform values, kept consistent across calls, instead of
querying $G$. The two games differ only if $\mathcal{A}$ queries $G$ on $(b\|s)$ for a hidden seed $s$. Until that
event, each hidden seed is uniform in $\{0,1\}^{\lambda}$ given $\mathcal{A}$'s view: the root is sampled uniformly,
and the value of a hidden non-root node is a random-oracle output on an input that $\mathcal{A}$ has not queried
(by induction from the root). A union bound over the $q_G$ queries gives
\[|\Pr[H_0]-\Pr[H_1]|\le q_G\,|\Phi|\,(n+1)/2^{\lambda}.\]
In $H_1$ the leaf value $r_t=\Fk(S,t)$ of every $t\in\Phi$ is uniform and independent of $\mathcal{A}$'s view.

\paragraph{$H_2$: closure keys become uniform}
Order the keys derived during the game by creation time. In $H_2$, every object key $k_o$ with $o\in\Cl(\Phi)$ and
every shard key whose shard version contains a member of $\Cl(\Phi)$ is replaced by a fresh uniform string, used
consistently. We argue by induction that each replaced key is the HKDF output on an input $x$ that contains a block
which is uniform and hidden:
\begin{itemize}
\item a source $o\in\Cl(\Phi)$ is in $\Phi$ (sources have no ancestors), and $k_o=r_{t_o}$ is uniform by $H_1$;
\item a derived $o\in\Phi$ has the block $r_{t_o}$ in its HKDF input;
\item a derived $o\in\Cl(\Phi)\setminus\Phi$ has a parent $p\in\Cl(\Phi)$ created earlier, whose key is uniform by
the induction hypothesis and appears as a block of $x$;
\item a shard key mixes the key of a member in $\Cl(\Phi)$.
\end{itemize}
HKDF inputs are pairwise distinct, because the \texttt{info} field contains the unique tag (objects) or the unique
shard identifier and version (shards). Hence each replaced output is a random-oracle answer on a fresh input, and
$H_1$ and $H_2$ differ only if $\mathcal{A}$ queries HKDF on one of these $N_K$ inputs, which requires guessing a
uniform 256-bit block:
\[|\Pr[H_1]-\Pr[H_2]|\le q_H\,N_K/2^{\lambda}.\]

\paragraph{$H_3$: challenge ciphertexts become independent of $b$}
In $H_2$ every ciphertext that encrypts challenge content is produced under a key that is uniform, independent of
everything else, and used for exactly one encryption: an object key encrypts one object once (a \textsc{Rotate}
derives new keys), and a shard key encrypts one shard version once. Validity guarantees that every challenge object
is in $\Cl(\Phi)$, hence covered. In $H_3$ we replace, one by one, each of the $C$ such ciphertexts by an encryption
of $0^{|m|}$ under the same key. Each step is bounded by the one-time IND-CPA advantage of an adversary $\mathcal{B}$
that simulates the rest of the game (it samples all other keys itself), so
\[|\Pr[H_2]-\Pr[H_3]|\le C\cdot\mathrm{Adv}^{\mathrm{ot\text{-}ind}}_{\mathrm{AEAD}}(\mathcal{B}).\]
In $H_3$ the view of $\mathcal{A}$ is independent of $b$ and $\Pr[H_3]=\tfrac12$. Summing the three bounds and
multiplying by two gives the theorem. \hfill$\blacksquare$

\paragraph{On re-derivation} Re-derived objects are ordinary new objects whose parents are surviving objects. They
are outside the closure of past forget requests, and the theorem says nothing about them; \cref{lem:confine} says
that, under the sound rule, their content does not depend on the forgotten sources.
\section{A Standard-Model Proof}\label{app:std}

We now remove the random oracles. Two changes are needed. First, the GGM PPRF is only known to be secure when the
punctured point is fixed in advance (\emph{selective} security); we avoid exponential guessing by guessing
\emph{one} forgotten ancestor per challenge object, at a polynomial cost. Second, HKDF must behave as a PRF when it
is keyed through one block of its input.

\begin{definition}[Block-keyed PRF]\label{def:bkprf}
For a position $j$, let $H^{(j)}_k(x_1,\dots,x_{m},\mathit{info})=\mathrm{HKDF}(x_1\|\cdots\|x_{j-1}\|k\|x_{j}\|\cdots\|x_m;\ \mathit{info})$
with $k\in\{0,1\}^{256}$. HKDF is a \emph{block-keyed PRF} if, for every $j$, $H^{(j)}$ is a PRF keyed by $k$,
with all other blocks and the info field chosen by the distinguisher. We write
$\mathrm{Adv}^{\mathrm{prf}}_{H}$ for the best advantage over all positions.
\end{definition}
With an empty salt, HKDF-Extract computes $\mathrm{HMAC}(0^{256},\cdot)$, so the definition asks HMAC to be a PRF
when keyed through its message. This is a dual-PRF-style assumption on the underlying compression function, of the
kind studied by Bellare \cite{bellare2015hmac}.

\paragraph{Key separation} In the random-oracle proof, an object key may be used both to encrypt the object and to
derive its children's keys. In the standard model we use the usual separation: the AEAD key of $o$ is
$\mathrm{HKDF}(k_o;\ \texttt{"lf/enc"})$ and only $k_o$ enters the children's derivations. This costs one HKDF call per
object; our implementation and measurements use $k_o$ directly, which the random-oracle theorem covers.

\begin{theorem}[LC-FS in the standard model]\label{thm:std}
Let $\mathcal{A}$ be a valid adversary that creates at most $N$ objects, of which $C'$ are challenge objects, and
let $\ell$ be the length of the longest path in the lineage graph and $V$ the largest number of shard versions that
contain one object. Then there are adversaries $\mathcal{B}_1,\mathcal{B}_2,\mathcal{B}_3$, running in about the
time of $\mathcal{A}$, such that
\begin{multline*}
\mathrm{Adv}^{\mathrm{lcfs}}_{\sys}(\mathcal{A})\le 2C'N\Big(n\cdot\mathrm{Adv}^{\mathrm{prg}}_{G}(\mathcal{B}_1)
+\ell\cdot\mathrm{Adv}^{\mathrm{prf}}_{H}(\mathcal{B}_2)\\
+(1+V)\cdot\mathrm{Adv}^{\mathrm{ot\text{-}ind}}_{\mathrm{AEAD}}(\mathcal{B}_3)\Big).
\end{multline*}
\end{theorem}

\begin{proof}
\emph{Step 0: one challenge at a time.} A standard hybrid over the $C'$ challenge objects, switching their content
from $m_0$ to $m_1$ one by one, reduces the game to adversaries with a single challenge object $d^\ast$, at a loss of
a factor $C'$.

\emph{Step 1: guess one forgotten ancestor.} Since $\mathcal{A}$ is valid, $d^\ast\in\Cl(\Phi)$, so there is an object
$a^\ast\in\Phi$ with $a^\ast=d^\ast$ or $a^\ast\in\Anc(d^\ast)$, and a path $a^\ast=u_0\to u_1\to\cdots\to u_L=d^\ast$
in the lineage graph with $L\le\ell$. The reduction guesses the index of $a^\ast$ among the $N$ objects that
$\mathcal{A}$ will create, and aborts if $a^\ast$ turns out not to be forgotten or not to be an ancestor of $d^\ast$.
The guess is correct with probability at least $1/N$ and is independent of $\mathcal{A}$'s view, which costs a factor $N$.

\emph{Step 2: the leaf of $a^\ast$ is pseudorandom.} Tags are drawn by the store, so the reduction picks
$x^\ast=t_{a^\ast}$ itself before any key exists. It obtains from the selective PPRF challenger the key punctured at
$x^\ast$ and a value $y^\ast$ that is either $F(S,x^\ast)$ or uniform. With the punctured key it evaluates every other
tag, uses $y^\ast$ for $x^\ast$, and applies the adversary's later forget requests to the punctured key. Since
$a^\ast\in\Phi$, the key it returns at corruption is exactly the one of the real game. Selective security of GGM
reduces to the PRG with a loss of $n$, one hybrid per tree level
\cite{boneh2013constrained,boyle2014functional,kiayias2013delegatable}.

\emph{Step 3: the path keys are pseudorandom.} With $y^\ast$ uniform, $k_{u_0}$ is uniform: it is $y^\ast$ if $u_0$ is a
source, and $H^{(j)}_{y^\ast}(\cdot)$ otherwise. For $i=1,\dots,L$, $u_{i-1}$ is a parent of $u_i$, so every
derivation that uses $k_{u_{i-1}}$ is an evaluation of $H^{(j)}$ keyed by $k_{u_{i-1}}$ on distinct inputs (the info
field contains the child's tag). Replacing these evaluations by a random function, one path step at a time,
makes $k_{u_i}$ uniform; this costs at most $\ell$ PRF advantages. The same argument covers the shard keys of the
versions that contain $d^\ast$, which are keyed through $k_{d^\ast}$.

\emph{Step 4: encryption.} The AEAD key of $d^\ast$ and the keys of the at most $V$ shard versions that hold its
vector are now uniform and each is used once, so the $1+V$ ciphertexts that depend on the challenge bit can be
switched by one-time IND-CPA.

Collecting the factors $C'$, $N$ and the factor 2 that turns probabilities into advantages gives the bound.
\end{proof}
The loss $C'N\ell$ is polynomial. For the random 32-bit tags of our implementation, the store must also avoid tag
collisions, which it does by rejection; a deployment that expects more than $2^{20}$ objects per key should use
64-bit tags, which only lengthens the GGM paths ($n=64$).
\end{document}